\documentclass[%
reprint,
superscriptaddress,
nofootinbib,
amsmath,amssymb,
aps,
prb,
]{revtex4-2}

\usepackage{graphicx}


\usepackage{bm}
\usepackage[colorlinks, linkcolor=blue, citecolor=blue, urlcolor=blue]{hyperref}

\DeclareMathOperator{\Tr}{Tr}
\DeclareMathOperator{\csch}{csch}
\DeclareMathOperator{\sech}{sech}
\DeclareMathOperator{\arctanh}{arctanh}
\newcommand{\norm}[1]{\left\lVert#1\right\rVert}

\usepackage{ulem}

\begin{document}

\title{
	A critical state under weak measurement is not critical
}

\author{Qicheng Tang}
\email{qtang73@gatech.edu}
\affiliation{School of Physics, Georgia Institute of Technology, Atlanta, GA 30332, USA}

\author{Xueda Wen}
\email{xueda.wen@physics.gatech.edu}
\affiliation{School of Physics, Georgia Institute of Technology, Atlanta, GA 30332, USA}

\date{\today}

\begin{abstract}
Critical systems host nontrivial entanglement structure that is generally sensitive to additional couplings. 
In the present work, we study the effect of weak measurements on the entanglement Hamiltonian of massless free fermions which are prepared in their critical ground state. 
While the power-law decaying correlation and logarithmic growing entanglement entropy have been observed as typical signatures of quantum criticality after the weak measurement~\cite{Altman2022_measure_critical, JianCM2023_measure_ising, Alicea2023_measure_ising, JianSK2023_measure_LL, Alicea2024_teleport_critical}, in this work we show that the conformal symmetry is lost and the entanglement Hamiltonian generally becomes gapped for arbitrary small measurement strength. 
To reveal this unconventional entanglement structure, we consider a field-theory description that allows us to establish an analytic mapping between the entanglement Hamiltonians before and after the weak measurements. 
From this mapping, we find that although the measurements lead to a significant modification of the entanglement spectrum, the real-space distribution of the eigenfunction of the kernel of entanglement Hamiltonian is unchanged, which is responsible for the coexistence of a gapped entanglement Hamiltonian and the logarithmic entanglement entropy. 
Moreover, as the mapping works for arbitrarily many disjoint intervals, the multi-interval entanglement entropy also exhibits the same scaling behavior as the critical ground state, and shares the same effective central charge with the single-interval case. 
We numerically demonstrate these field-theory results on a lattice model, where the entanglement Hamiltonian after weak measurements exhibits the typical signature of a finite gap and becomes long-ranged even in the single-interval case. This is distinct from the critical ground state, where the entanglement Hamiltonian for a single interval is gapless and local. 
\end{abstract}

\maketitle


\section{Introduction}

The concept of entanglement is fundamental in modern physics and powerful for investigating the intrinsic 
structure of many-body systems~\cite{Amico2007_rmp, Laflorencie2015_review, Wen2017_zoo}. 
Well-established examples including probing universality of critical systems~\cite{Wilczek1994_entropy_cft, Vidal2002_EE_critical, Calabrese2004_EE_QFT, Wolf2006_EE_fermion, Klich2006_EE_fermion, Cardy2016_EH_cft}, identifying topological order~\cite{KitaevPreskill2005_topoEE, LevinWen2006, LiHaldane2008, Pollmann2010_ES_topo, Qi2012_ES_topo_quench}, characterizing non-equilibrium phases of matter~\cite{Prosen2008_MBL, Pollmann2012_unbound_EE_MBL, Serbyn2013_growEE_MBL, Skinner2018_MIPT, Li2018_MIPT}, and so on. 
Among those applications, the scaling behavior of entanglement entropy (EE) -- a typical measure of entanglement, has attracted particular attention. 
As predicted by conformal field theory (CFT)~\cite{CFT_yellow_book}, $(1+1)$-dimensional critical systems host a sub-extensive entanglement structure with logarithmic growing EE~\cite{Wilczek1994_entropy_cft, Vidal2002_EE_critical, Calabrese2004_EE_QFT}. 
This behavior is distinct from the area law applicable to general low-energy states of local theories~\cite{Bombelli1986_area, Srednicki1993_area, Eisert2008_rmp} and the volume law associated with high-energy excitations or systems out of equilibrium~\cite{Page1993, Srednicki1994_ETH, Calabrese2005_quench_EE, Alba2009_EE_excited, Rigol2017_EE_quadratic}, making it a typical signature of criticality.

A problem of interest from both theoretical and experimental perspectives is the effect of additional couplings with the environment. Unlike ordinary local observables, entanglement in many-body systems is particularly sensitive to these additional couplings. One might wonder whether the nontrivial sub-extensive entanglement structure of critical systems can survive under certain measurement protocols. In the context of CFT, the \textit{weak measurement} can be viewed as an interface separating the pre- and post-measurement states~\cite{Rajabpour2015_measurement_cft, Watanabe2016_project_holography, Swingle2022_holo_measure, Popov2022_MIPT_holo, Swingle2022_holo_measure_2, Swingle2023_holo_measure, JianCM2023_measure_ising, Alicea2023_measure_ising, JianSK_2023_holo_weak_measure, JianSK2023_measure_LL, Wei2023_MIPT_holo, Alicea2024_teleport_critical}, and is related to an interface CFT after a spacetime rotation~\cite{JianCM2023_measure_ising, JianSK_2023_holo_weak_measure}. 
This approach leads to the analytical prediction of logarithmic EE scaling (with respect to the subsystem size) with a continuously tunable effective central charge after weak measurements, when the interface is marginal to the bulk CFT~\cite{JianCM2023_measure_ising, Alicea2023_measure_ising, JianSK_2023_holo_weak_measure, JianSK2023_measure_LL, Alicea2024_teleport_critical}. 
Based on this consideration, the post-measurement state is considered to be ``altered'' by measurements and remains to be critical~\cite{Altman2022_measure_critical, JianCM2023_measure_ising, Alicea2023_measure_ising, JianSK2023_measure_LL, Alicea2024_teleport_critical}.

In the present work, we explore the detailed entanglement structure of the critical ground state under weak measurements. In particular, we consider the ground state of critical Dirac fermions, with a measurement protocol that leads to a Néel-order-like measurement outcome, resulting in a scale-invariant post-measurement state with power-law decaying correlation function and logarithmically growing EE. 
Since these scaling behaviors are typical signatures of criticality, one might naively think that the post-measurement state is critical. However, through an analytical calculation in the continuum limit, we show that these measurement-induced ``critical'' states are, in fact, non-critical, as they have a gapped entanglement Hamiltonian (EH). 
In addition, we find that the post-measurement states behave quite differently from a CFT ground state and host an EH with long-range couplings, which is intrinsic to the coexistence of a gapped entanglement spectrum (ES) and logarithmic EE.

The present paper is organized as the follows. 
In Sec.~\ref{sec:model}, we introduce the lattice model of weak measurements for free Dirac fermions and the corresponding field-theory description at continuum. 
To investigate the entanglement structure of the state, we discuss on the spectral decomposition of the correlation matrix in Sec.~\ref{sec:spectra_correlation}. This leads to an exact mapping of the EH before and after measurements as discussed in Sec.~\ref{sec:exact_ES_EE} \&~\ref{sec:field_theory_EH}. 
The analytical results of ES  and EE are then demonstrated via numerical simulations on the lattice model, which also provide more detailed information about the EH after measurements.

\section{Weak measurements for free Dirac fermion}\label{sec:model}

In this work, we investigate a model of applying weak measurements on the ground state of $(1+1)$-dimensional free Dirac fermions, with a N\'eel-order-like measurement outcome of the physical system along the occupation number basis. 
Under a weak measurement limit, it can be converted to the following imaginary time evolution~\cite{Altman2022_measure_critical, JianCM2023_measure_ising, Alicea2023_measure_ising, JianSK2023_measure_LL, Alicea2024_teleport_critical}
\begin{equation}\label{eq:imag_time_evo_lattice}
| \psi_{M} \rangle = \frac{e^{-\beta H_{M} } |\psi_{\rm GS} \rangle}{|| e^{-\beta H_{M} } |\psi_{\rm GS} \rangle ||} , \ 
H_{M} = \sum_{m=1}^{L} (-1)^m c_m^\dagger c_m 
\end{equation}
where $\beta\ge 0$ characterizes the strength of the weak measurement,
and $|\psi_{\rm GS} \rangle$ is the ground state of 
\begin{equation}\label{eq:massless_fermion_chain}
H_0 = - \frac{1}{2} \sum_{m=1}^{L} c_m^\dagger c_{m+1} + h.c., 
\end{equation}
where the fermionic operators $c_m$ satisfy the anticommutation relation 
$\{c_m,c_n^\dag\}=\delta_{mn}$, and $\{c_m, c_n\}=\{c_m^\dag, c_n^\dag\}=0$. 
Here we are interested in the half-filling case with periodic boundary conditions. By performing Fourier transformation, we have the following two-point correlation function after weak measurements
\begin{equation}\label{eq:correlator_lattice}
\langle c_m^\dagger c_n \rangle_\beta = 
\begin{cases}
\frac{1}{2} \frac{ e^{-2\beta} }{ \cosh 2\beta }
\quad & {\rm even} \ n = m,  \\ 
\frac{1}{2} \frac{ e^{2\beta} }{ \cosh 2\beta }
\quad & {\rm odd} \ n = m, \\ 
0  \quad & {\rm even} \ n - m \neq 0 , \\ 
\frac{1}{\cosh 2\beta}
\frac{i}{\pi} (n - m)^{-1} \ &  {\rm odd} \ n - m \neq 0 , \\ 
\end{cases}
\end{equation}
where we have taken the thermodynamic limit $L \to \infty$ (see details in Appendix~\ref{app:lattice_solution}). 

A field-theory description at continuum can be constructed as follows. 
First note that the imaginary time evolution in Eq.~\eqref{eq:imag_time_evo_lattice} effectively couples the chiral and anti-chiral modes of free massless Dirac fermions in a way of $H_M = \int \overline{\Psi}(x) \Psi(x) dx$, where $\overline{\Psi}(x, 0) = \Psi^\dagger(x, 0) \gamma^0$. 
This leads to the following equation of motion of the Dirac field under weak measurements 
\begin{equation}
(\gamma^0 \partial_\tau + \beta_0) \Psi(x, \tau) = 0 ,
\end{equation}
with an initial condition that the two-component spinor $\Psi(x, 0) =\left( \Psi_1(x,0), \Psi_2(x,0) \right)^T$ is the ordinary solution of free massless Dirac field (see Appendix~\ref{app:field_solution} for details). 
Here $\beta_0$ is the unit scale of imaginary time, for simplicity we let $\beta_0 = 1$. 
This leads to the following two-point correlation function
\begin{equation}\label{eq:two_point_field}
\begin{aligned}
& C(x, y; \beta) = \langle 0 | \Psi(x, \beta) \Psi^\dagger(y, \beta) | 0 \rangle \\
= & \frac{1}{2} (1 - \gamma^0\tanh 2\beta ) \delta(x - y) 
- \frac{i}{2\pi} \frac{1}{\cosh 2\beta} \frac{\gamma^1 \gamma^0}{x - y} ,
\end{aligned}
\end{equation}
which is consistent with the lattice result in Eq.~\eqref{eq:correlator_lattice}, but with open boundary conditions. For the Dirac matrices, we choose $\gamma^0 = \sigma_2, \gamma^1 = i\sigma_1$, and $\gamma^0 \gamma^1 = \sigma_3$. 

An important observation here is that the power-law decaying correlation function remains with the same critical exponent, for which a similar result was reported in the Ising model with measurements along transverse direction~\cite{Alicea2023_measure_ising, JianCM2023_measure_ising, Alicea2024_teleport_critical} and the Luttinger liquid with a charge-density-wave-like measurement outcome~\cite{Altman2022_measure_critical, JianSK2023_measure_LL}.
The presence of a power-law correlation indicates the existence of scale invariance as the pre-measurement critical ground states, but the short-distance behavior is strongly modified to host an even-odd effect in the diagonal terms of correlation matrix. 
This N\'eel-order-like behavior implies a gap opening for the EH after weak measurements. 
Later we will discuss these structure in detail.

\section{The spectral information of subsystem correlation matrix}\label{sec:spectra_correlation}

As we are considering a Gaussian theory, the reduced density matrix can be fully addressed in terms of two-point functions. In particular, the EH of free fermions can be written as
\begin{equation}\label{eq:def_EH_kernel}
K_A = - \frac{1}{2\pi} 
\log \rho_A = \int_A dx \int_A dy : \Psi^\dagger(x) \mathcal{H}_A(x, y) \Psi(y) :,  
\end{equation}
with the kernel~\cite{Chung2001_correlation_matrix, Peschel2003_correlation_matrix}
\begin{equation}\label{eq:connect_EH_CM}
\mathcal{H}_A(x, y) = - \frac{1}{2\pi} 
\log (C_A^{-1}(x, y) - 1) ,
\end{equation}
where $C_{A}(x,y)$ is the subsystem correlation matrix for $x, y \in A$ and $: :$ represents normal ordering. 
Here the subsystem can be either a single interval $A = (a, b)$ or a set of multiple disjoint intervals $A = A_1 \cup \dots A_N = (a_1, b_1) \cup \dots (a_N, b_N)$ with $a_i < b_i < a_{i+1}$. 
To analytically calculate the (kernel of) EH after weak measurements, we need to solve the following spectral decomposition
\begin{equation}\label{eq:spectral_decomp_CA}
\int_A C(x, y; \beta) \Phi_s(y) dy = \eta_s \Phi_s(x) ,
\end{equation}
where $s$ is a good quantum number to label the eigenvectors. 
Notice that there are only two operators $\frac{1}{x - y}$ and $\delta(x-y)$ in the correlation matrix $C_A(x, y; \beta)$, just as the ground state case. 
The only difference is that the chiral and anti-chiral mode are now coupled in terms of a Dirac-$\delta$ function.
It is then natural to consider that the correlation matrix after weak measurements shares the same orthogonal basis with the ground state case. 

For the single-interval case, the EH of the later can be diagonalized by a conformal mapping to an annulus~\cite{CardyTonni2016_EH}, where the plane wave forms a complete set of orthogonal basis. 
Combine with Eq.~\eqref{eq:connect_EH_CM}, it leads to the following spectral decomposition~\cite{CasiniHuerta2009_fermion_EH, Casini2018_chiral_scalar, Hollands_2019, Tonni2020_EH_Dirac_boundary}
\begin{equation}\label{eq:eigen_singular_correlation_op}
\int_A \frac{1}{x - y} \phi_{s_0}(y) dy = \lambda_{s_0} \phi_{s_0}(x) 
\end{equation}
with the eigenvalue
\begin{equation}
\lambda_{s_0} = \pi i \tanh \pi s_0
\end{equation}
where $s_0\in\mathbb R$,
and the eigenvector
\begin{equation}\label{eq:pure_eigenvector}
\phi_{s_0}(x) 
= \sqrt{ \frac{ z'(x) }{2\pi} } e^{-i s_0 z(x)} , 
\ 
z(x) = \log \frac{x - a}{b - x} .
\end{equation}
Here $z(x)$ is the conformal mapping from cutting a single interval on the plane to an annulus, 
where the EH is the translation generator with eigenvalue $s_0$~\footnote{In the context of CFT, it is well known that the EH on an annulus is proportional to the physical CFT Hamiltonian~\cite{CardyTonni2016_EH}, and therefore shares the same eigenvector. For the free Dirac fermions, the eigenvectors of a physical Hamiltonian are simply the plane waves, but now on the annulus after conformal mapping. Combining with the fact that the EH is a function of correlation matrix as shown in Eq.~\eqref{eq:connect_EH_CM} (again, this is a consequence of Gaussianality of free Dirac fermions), we have $e^{- i s_0 z(x)}$ as the guess solution of Eq.~\eqref{eq:eigen_singular_correlation_op}. By imposing the orthogonal and complete conditions of eigenvectors, it is then straightforward to obtain Eq.~\eqref{eq:pure_eigenvector}.}. 

The multi-interval cases are more complicated to handle, since there is generally no local expression of EH in terms of a conformal mapping~\cite{CasiniHuerta2009_fermion_EH, Casini2018_chiral_scalar, Eisler2022_TwoInt}. 
Fortunately, the spectral decomposition problem of Eq.~\eqref{eq:eigen_singular_correlation_op} is solvable by considering a generalized mapping of
\begin{equation}\label{eq:pure_eigenvector_multi}
z(x) = \log \left( \prod_{i=1}^{N} \frac{x - a_i}{b_i - x} \right) ,
\end{equation}
which leads to the following eigenvector for multi-interval~\cite{CasiniHuerta2009_fermion_EH, Casini2018_chiral_scalar}
\begin{equation}
\phi_{s_0}^{k}(x) = \frac{\Theta(x)}{\mathcal{N}_k} 
\frac{ \prod_{i \neq k} (x - a_i) }{\sqrt{ \prod_{i=1}^{N} |x - a_i| |x- b_i| }} 
e^{-i s_0 z(x)} , 
\end{equation}
where $\Theta(x) = (-1)^{j+1}$ for $x \in (a_j, b_j)$ in $j$-th interval. 
Here $k = 1, \dots, N$ labels the degenerate eigenvectors with eigenvalue $\lambda_{s_0}$ that are contributed by the multiple intervals, and $\mathcal{N}_k = \sqrt{ 2\pi { \prod_{i \neq k} (a_i - a_k) } / { \prod_{i=1}^{N} (b_i - a_k) } }$ is the normalization factor. 
As we take the number of disjoint intervals $N = 1$, this solution reduces to the single-interval case shown in Eq.~\eqref{eq:pure_eigenvector}. 
Moreover, it satisfies 
\begin{equation}\label{eq:product_eig_trace}
\sum_{k = 1}^{N} \phi_{s_0}^{k} (x) \phi_{s_0}^{k *} (x)  
= \frac{1}{2\pi} \sum_{i = 1}^{N} \left( \frac{1}{x - a_i} - \frac{1}{x - b_i} \right)  , 
\end{equation}
which is nothing but $\frac{1}{2\pi} z'(x)$. 
This relation is very useful for further evaluation of the EE.

Before discussing the effect of weak measurement, it is worth to revisit the ground state of massless Dirac fermions in advance, for which we have the following eigenvalues 
\begin{equation}
\eta_{s_0} 
= \left( \begin{matrix}
\eta_{s_0}^{+} & 0 \\ 0 & \eta_{s_0}^{-}
\end{matrix} \right)
= \left( \begin{matrix}
\frac{1}{2} + \frac{i}{2\pi} \lambda_{s_0} & 0 \\
0 & \frac{1}{2} - \frac{i}{2\pi} \lambda_{s_0}
\end{matrix} \right) 
\end{equation}
and eigenvectors
\begin{equation}
\Phi_{s_0}(x) = 
\left( \begin{matrix}
\Phi_{s_0}^{+}(x) & \Phi_{s_0}^{-}(x) 
\end{matrix} \right)
= 
\left( \begin{matrix}
\phi_{s_0}(x) & 0 \\ 0 & \phi_{s_0}(x)
\end{matrix} \right) . 
\end{equation}
Here the different signs $\{ + , - \}$ of the eigenvalues and eigenvectors correspond to the chiral and anti-chiral modes, respectively. 
Now, let us introduce the ansatz for solving Eq.~\eqref{eq:spectral_decomp_CA} after weak measurements: the new eigenfunction is a linear combination of the ground-state solution as $\Phi_s = c_1 \Phi_{s_0}^{+} + c_2 \Phi_{s_0}^{-}$. 
After some straightforward algebra, this leads to the following eigenvalues of the correlation matrix after weak measurements
\begin{equation}\label{eq:eigenvalue_CM}
\eta_s^{\pm} 
= \frac{1}{2} \pm \frac{1}{2} 
\sqrt{ \left( \frac{i \lambda_{s_0}}{\pi \cosh 2\beta} \right)^2 + \left( \tanh 2\beta \right)^2 }
, 
\end{equation}
and the eigenvectors
\begin{equation}\label{eq:eigenvector_CM}
\Phi_s(x) 
= \left( \begin{matrix} \Phi_s^+(x) & \Phi_s^-(x) \end{matrix} \right) 
= \frac{1}{ \mathcal{N}_\beta } \left(\begin{matrix}
\phi_{s_0}(x) &  -t \phi_{s_0}(x) 
\\ 
-t \phi_{s_0}(x) & \phi_{s_0}(x) 
\end{matrix}\right) , 
\end{equation}
where $|\mathcal{N}_\beta|^{2} =  (|t|^2+1) = (-t^2 + 1)$ is the normalization factor and $t = \frac{ \csch 2\beta}{\pi} \lambda_{s_0} + \sqrt{ \left( \frac{ \csch 2\beta}{\pi} \lambda_{s_0} \right)^2 - 1 }$ is a pure imaginary function of $s_0$ and $\beta$. 
We can also write the eigenvectors in Eq.~\eqref{eq:eigenvector_CM} in terms of
\begin{equation}
\label{Unitary_beta}
\Phi_s(x) = U_s(\beta) \Phi_{s_0}(x) 
= \frac{1}{\mathcal{N}_\beta} \left( \begin{matrix}
1 & -t \\ -t & 1
\end{matrix} \right) \Phi_{s_0}(x) . 
\end{equation}
Here $U_s(\beta)$ is a unitary transformation, so that the new eigenvectors still form a complete set of orthogonal basis to satisfy the orthogonality condition 
$\int_A dx\, \Phi_s(x) \Phi_{s'}^\dagger(x) = \delta(s - s') \mathbb{I}_{2 \times 2}$
and the completeness condition 
$\int ds\, \Phi_s(x) \Phi_s^\dagger(y) = \delta(x - y)  \mathbb{I}_{2 \times 2}$. 

The unitary transformation $U_s(\beta)$ in Eq.~\eqref{Unitary_beta} does not rely on the configuration of the subsystem, therefore we expect universal results for both single-interval and multi-interval cases. 
In fact, as we will address in the following sections, the ES transforms in a universal way and leads to a single effective central charge for both single-interval and multi-interval EE. 
Moreover, it should be noticed that $U_s(\beta)$ is a nontrivial function of the eigenvalue $s_0$ and singular at $\beta = 0$. 
As a consequence, this solution is only valid for finite $\beta > 0$, but cannot be perturbatively connected to the ground state case. 
This fact indicates a sudden change of the EH for arbitrary nonzero measurement strength, which we will discuss in detail later.

\section{Entanglement spectrum and entropy after weak measurements}\label{sec:exact_ES_EE}

From Eq.~\eqref{eq:eigenvalue_CM} we have the following relation for both single-interval and multi-interval cases~\footnote{ Here we notice that the same relation was obtained for the interface CFT to establish the mapping with and without conformal interface for a critical free fermionic chain~\cite{Peschel2010_EE_defect}, but in a different form of
$ \cosh \pi s = \cosh 2\beta \cosh \pi s_0 $. 
In the case of interface CFT, the mapping is only valid for a subsystem with one end at the physical boundary and another at the interface. However, here we have this relation for arbitrary subsystems, including multi-interval cases.
This difference comes from the fact that our setup is translationally invariant in space, while the setup of interface CFT breaks the translation symmetry due to the insertion of conformal interface at a specific location.}
\begin{equation}\label{eq:entanglement_spectrum}
\tanh^2 \pi s = \left( \frac{\tanh \pi s_0}{\cosh 2\beta} \right)^2 + \left( \tanh 2\beta \right)^2 ,
\end{equation}
where we have used $\eta_s^{\pm} = \frac{1}{2} \pm \frac{1}{2} \tanh \pi s$ from the definition of Eq.~\eqref{eq:connect_EH_CM}. 
Here $s_0$ and $s$ are ES before and after the weak measurements, respectively. 
As the ground state ES $s_0$ is gapless, Eq.~\eqref{eq:entanglement_spectrum} indicates a finite gap in the ES after weak measurements 
\begin{equation}\label{eq:entanglement_spectrum_gap}
\Delta = \frac{4}{\pi} \beta 
\end{equation}
that is linear in the measurement strength $\beta$. 

We further calculate the $n$-th R\'enyi entropy after a weak measurement:
\begin{equation}
\begin{aligned}
S_{A}^{(n)} & = \Tr \left[ g^{(n)} (C_A) \right]
\\ & 
= \int_{A} dx \int ds \Tr \left[ \Phi_{s}(x)  g^{(n)} (\eta_s)  \Phi_{s}^{\dagger}(x) \right], 
\end{aligned}
\end{equation}
where $g^{(1)}(u) = - u \log u - (1 - u) \log (1 - u)$, 
$g^{(n \neq 1)}(u) = \frac{1}{1-n} \log \left[ u^n + (1-u)^n \right]$. 
It is important to notice that the two symmetric eigenvalues $\eta_{s}^{+}$ and $\eta_{s}^{-}$ give the same value of $g^{(n)}(\eta_s)$. 
More importantly, the trace of eigenvector 
\begin{equation}\label{eq:trace_invariant}
P_s(x) = \Tr \left[ \Phi_{s}(x) \Phi_{s}^{\dagger}(x) \right] 
= \Tr \left[ \Phi_{s_0}(x) \Phi_{s_0}^{\dagger}(x) \right] 
\end{equation}
is the same as that before weak measurements and is independent of $s_0$. 
These lead to the following separation of the R\'enyi entropy as
\begin{equation}
S_A^{(n)} = \Bigg[ \int_A dx P_s(x)  \Bigg] \Bigg[ \int ds g^{(n)}(\eta_s) \Bigg] . 
\end{equation}
Here the first part is an integral of $P_s(x)$ over the real space. As $P_s(x)$ remains unchanged, this integral gives the same scaling behavior as the ground state. 
In particular, from the relation in Eq.~\eqref{eq:product_eig_trace}, we have
\begin{equation}
\begin{aligned}
\int_{A} dx P_s(x) 
& = 2 \int_A dx \, \phi_{s_0}(x) \phi_{s_0}^{*}(x)
\\ & = \frac{1}{\pi} \int_{A} dx \sum_{i=1}^{N} 
\left( \frac{1}{x - a_i} -  \frac{1}{x - b_i} \right) .
\end{aligned}
\end{equation}
For the single-interval case, it gives
\begin{equation}
\begin{aligned}
\int_A dx P_s(x) 
= \int_{A_\epsilon} dx P_s(x) 
= \frac{2}{\pi} \log \frac{l}{\epsilon}, 
\end{aligned}
\end{equation}
where $l = b - a$ is the subsystem size and we have introduced a regularization of $A = (a, b) \to A_\epsilon = (a + \epsilon, b - \epsilon)$. 
Similarly we can also obtain the scaling behavior of two (or more) intervals as
\begin{equation}\label{eq:EE_scaling_TwoInt}
\int_{A_{1, \epsilon} \cup A_{2, \epsilon}} dx P_s(x) 
= \frac{2}{\pi} \log \frac{ l_1 l_2 r (l_1 + l_2 + r) }{\epsilon^2 (l_1 + r) (l_2 + r)} ,
\end{equation}
where $l_1 = b_1 - a_1$ and $l_2 = b_2 - a_2$ are the subsystem sizes of $A_1$ and $A_2$, $r = a_1 - b_2$ is the distance between them. 
As the logarithmic scaling of EE remains, the effect of weak measurements only leads to a modification of its prefactor to be
\begin{equation}\label{eq:integral_c_eff}
\begin{aligned}
\sigma^{(n)}(\beta) = \frac{2}{\pi} \int ds \, g^{(n)}(\eta_s) 
= \frac{2}{\pi} \int_{-\infty}^{\infty} d s_0 \frac{ds}{ds_0} g^{(n)}(\eta_{s}) ,
\end{aligned}
\end{equation}
where $s$ and $\eta_s$ in terms of $s_0$ are given by Eq.~\eqref{eq:eigenvalue_CM} \&~\eqref{eq:entanglement_spectrum}. 
Although a general closed form is hard to obtain, it is easy to check that this expression gives the ordinary ground-state results by taking $\beta \to 0$, i.e.
\begin{equation}
\begin{aligned}
\sigma^{(n)}(0) & = \frac{2}{\pi} \frac{1}{1-n} \int_{-\infty}^{\infty} 
\log \left[ \sum_{\kappa = \pm 1} \left( \frac{1}{2} + \kappa \frac{1}{2} \tanh \pi s_0 \right)^n \right] d s_0 
\\ & = \frac{n+1}{6n} . 
\end{aligned}
\end{equation}
Moreover, one can show that the integral of Eq.~\eqref{eq:integral_c_eff} is identical to the previous results on critical Ising chain upto a global factor of $2$. 
At the replica limit $n \to 1$, the integral gives an effective central charge as a free-fermion interface CFT~\cite{Peschel2010_EE_defect, Peschel2012_EE_defect, Alicea2024_teleport_critical}
\begin{equation}\label{eq:defect_effective_central_charge}
\begin{aligned}
c_{\rm eff}^{(1)} = 3 \sigma^{(1)} = -\frac{6}{\pi^2} \sum_{\kappa = \pm 1}
\Big\{ (1+\kappa \theta) \log (1+ \kappa \theta) \log \theta \\ 
+ (1+\kappa \theta) {\rm Li}_2(-\kappa \theta)  \Big\} ,
\end{aligned}
\end{equation}
where $\theta = \frac{1}{\cosh 2\beta}$. This is consistent with the previous result of converting the weak measurements to conformal interface via a spacetime rotation of the path integral~\cite{JianCM2023_measure_ising, JianSK_2023_holo_weak_measure}.

\section{The structure of entanglement Hamiltonian: insights from the field theory at continuum}\label{sec:field_theory_EH}

We have shown that the ES after weak measurements becomes gapped, but the distribution function $P_s(x)$ is unchanged, resulting in a logarithmic scaling of EE similar to the ground state. 
In this section, we are going to reveal the coexistence of a gapped ES and a logarithmic EE by investigating more about the structure of EH, for which the kernel has the following integral representation
\begin{equation}\label{eq:int_rep_ker_EH}
\mathcal{H}_A(x,y) 
= - \int s \,\Phi_s(x) \sigma_3 \Phi_s^\dagger(y) ds .
\end{equation}
Before discussing the weak-measurement case, let us first briefly introduce the known results for ground states at half filling. 
From Eq.~\eqref{eq:pure_eigenvector} \& \eqref{eq:int_rep_ker_EH}, it is straightforward to obtain the kernel as
\begin{equation}
\begin{aligned}
\mathcal{H}_A^{\rm GS}(x, y) 
& = - \sigma_3 \frac{\sqrt{z'(x) z'(y)}}{2\pi} \int_{-\infty}^{\infty} s_0 e^{-s_0 \left( z(x) - z(y) \right)} ds_0 \\
& = - \frac{i \sigma_3}{2}  \sum_{y = x_i} \delta(x - y) \left[ - \frac{\partial_x}{z'(x)} + \frac{\partial_y}{z'(y)} \right] , 
\end{aligned}
\end{equation}
where the integral over $s_0$ in the first line is a Fourier transformation from $s_0$ to $z(x) - z(y)$, and $y = x_i$ in the last line are roots of $z(x) - z(y) = 0$. 
By the definition of Eq.~\eqref{eq:def_EH_kernel}, this leads to 
\begin{equation}
K_A^{\rm GS}= \int_A \frac{1}{z'(x)} \Psi^\dagger(x) (-i \sigma_3 
\partial_x) \Psi(x)  dx . 
\end{equation}
Recall that $z(x)$ is the conformal mapping from a plane to an annulus, it is then obvious that this is identical to the CFT solution of
\begin{equation}\label{eq:entanglement_Hamiltonian_CFT_GS}
K_A^{\rm GS} = \int_A \frac{1}{z'(x)} T_{00}(x) dx ,
\end{equation}
where $T_{00}(x) = \Psi^\dagger(x) (-i \gamma^0 \gamma^1 \partial_x) \Psi(x)$ is the CFT Hamiltonian density, with $\gamma^0 \gamma^1 = \sigma_3$. 
Since the EH is a local integral of $T_{00}(x)$, it is still gapless and local as the physical CFT Hamiltonian.

Now let us turn to considering the weak-measurement case. From Eq.~\eqref{eq:eigenvalue_CM} \&~\eqref{eq:eigenvector_CM}, we have the following explicit form of Eq.~\eqref{eq:int_rep_ker_EH}
\begin{equation}\label{eq:explicit_EH_weak_measure}
\begin{aligned}
& \mathcal{H}_A(x,y) = \int_{-\infty}^{\infty} F(s_0, \beta) \phi_{s_0}(x) \phi_{s_0}^{*}(y) ds_0 
\\ & = \int_{-\infty}^{\infty} F(s_0, \beta) e^{-is_0 ( z(x) - z(y) )} ds_0 \frac{\sqrt{z'(x) z'(y)}}{2\pi} ,
\end{aligned}
\end{equation}
with 
\begin{equation}
\begin{aligned}
F(s_0, \beta) & = \frac{1}{\pi} \arctanh
\sqrt{ \frac{\tanh^2 \pi s_0}{\cosh^2 2\beta} + \tanh^2 2\beta } 
\\ 
& \quad \times 
\frac{1}{(1 - t^2)} \left( \begin{matrix}
1+t^2 & 2t \\ -2t & -1-t^2 
\end{matrix} \right) , 
\end{aligned}
\end{equation}
where $t$ is a pure imaginary function of $s_0$ and $\beta$ as shown in Eq.~\eqref{eq:eigenvector_CM}.

Instead of an exact solution, here we consider a ``low-energy'' expansion of $F(s_0, \beta)$ around $s_0 \sim 0$, such that we simplify $F(s_0, \beta)$ to be a polynomial of $s_0$. 
For the diagonal terms, their Taylor expansion around $s_0 \sim 0$ gives only odd-order terms as 
\begin{equation}\label{eq:diag_EH_ker}
F_{11}(s_0, \beta) 
= - F_{22}(s_0, \beta) 
= \sum_{n=0}^{\infty} d_n(\beta) s_0^{2n+1} , 
\end{equation}
where the factor of lowest-order term $d_0(\beta) = 2\beta \csch 2\beta$ is a monotonically deceasing function with $\lim_{\beta \to 0} d_0 = 1$, and all higher-order terms of $d_n(\beta), n \ge 1$ are in the order of $\mathcal{O}(\beta^2)$. 

In contrast, the Taylor expansion of off-diagonal terms around $s_0 \sim 0$ only contain even-order terms as
\begin{equation}\label{eq:off_diag_EH_ker}
F_{12}(s_0, \beta) = F_{21}^*(s_0, \beta) = \sum_{n=0}^{\infty} h_n(\beta) s_0^{2n} ,
\end{equation}
with $h_0(\beta) = i\frac{2\beta}{\pi}$ for the lowest-order term. 
It is easy to show that all $h_n(\beta)$ are pure imaginary (since $t$ is pure imaginary) and in the order of $\mathcal{O}(\beta)$ so that the off-diagonal terms vanish at the limit $\beta \to 0$. 

\begin{figure*}\centering
	\includegraphics[width=\textwidth]{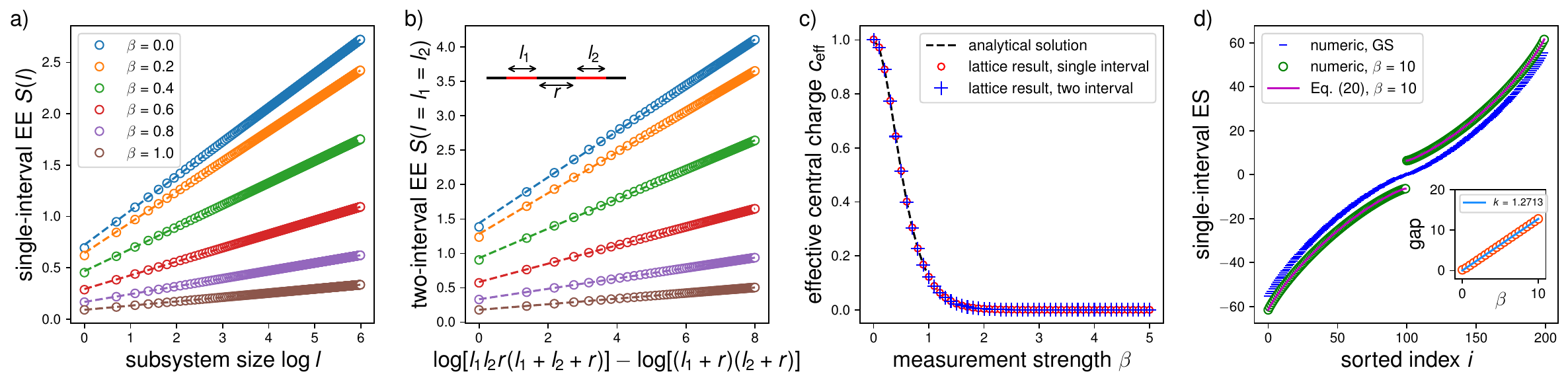}
	\caption{
		\label{fig:EE_ES}
		Coexistence of scale-invariant scaling behaviors of the EE and a gapped ES after weak measurements. 
		a) The single-interval EE as a logarithmic function of subsystem size $l$. 
		b) The two-interval EE with a varying subsystem size $l = l_1 = l_2$ and a fixed distance $r = 5$ between the two equal-length subsystems, plotted in the appreciated scaling form as shown in Eq.~\eqref{eq:EE_scaling_TwoInt}. 
		We have also checked that general configurations of the two intervals satisfy the same scaling form with respect to $l_1, l_2, r$. 
		c) The effective central charge extracted from the numerical data in a) and b), showing a perfect agreement with the analytical solution of Eq.~\eqref{eq:defect_effective_central_charge}. 
		d) The sorted single-interval ES with a fixed subsystem size $l = 200$, showing a perfect agreement with the analytical prediction of a relation of Eq.~\eqref{eq:entanglement_spectrum} between the ES of critical ground state and after weak measurements. The inset represents the dependence of gap $\Delta$ of ES on the measurement strength $\beta$, for which a linear fitting of $\Delta  = k \beta + b$ gives $k \approx 1.2713 \approx \frac{4}{\pi}$. 
		Here all numerical data are obtained by considering an infinite total system size, which has the exact two-point correlation function shown in Eq.~\eqref{eq:correlator_lattice}. 
	}
\end{figure*}

The above low-energy expansions allow us to construct an effective theory with (infinite) higher-order derivatives to analyze the structure of EH after weak measurements. 
To see this, consider the following Fourier transformation
\begin{equation}\label{eq:Fourier_power}
\begin{aligned}
\mathcal{I}_n 
& = \int_{-\infty}^{\infty} s_0^n e^{- i s_0 \left( z(x) - z(y) \right)} ds_0 
\\
& = \pi i^{n} \sum_{y = y_i} \frac{\delta(x - y)}{z'(x)} 
\left[ - \frac{\partial_x}{z'(x)}  + \frac{\partial_y}{z'(y)} \right]^n ,
\end{aligned}
\end{equation}
where $\delta^{(n)}$ represents $n$-th order derivative of the Dirac-$\delta$ function with $\delta^{(0)}(x-y) = \delta(x-y)$, and $y_i$ are the roots of $z(x) - z(y) = 0$.
For the single-interval case, we simply have a single root of $y_i = y$. 

By using Eq.~\eqref{eq:Fourier_power}, it is then straightforward to calculate the low-energy approximation of the EH after weak measurements. 
The higher-order derivatives imply the presence of long-range couplings in the EH after lattice discretization. 
Each term of $\mathcal{I}_n$ contributes multiple lower-order derivatives on the field operator with position-dependent weights to the EH, therefore it is hard to obtain a closed form. 
However, it is worth to notice that even-order terms of $\mathcal{I}_n$ only contribute even-order derivatives on the field operator, while odd-order terms of $\mathcal{I}_n$ only contribute odd-order derivatives. 
After a discretization to lattice models, the hopping terms with even distance will be dominated by the off-diagonal components of $K_A$, while odd-distance terms are dominated by diagonal components of $K_A$, so that they would behave very differently.

Among the emergence of (infinite) long-range couplings after weak measurements, we can further explore more detailed structure of the EH by looking at the leading terms of $n = 0$. 
For the diagonal terms we have
\begin{equation}\label{eq:NN_EH_measure}
K_{A, {\rm diag}}^{n=0}  = 2 \beta \csch 2\beta  \int_A 
\Psi^\dagger(x)  \frac{- i \sigma_3 \partial_x}{z'(x)}  \Psi(x) dx ,
\end{equation}
with a prefactor that monotonically deceases with $\beta$. 
For the off-diagonal terms we have
\begin{equation}\label{eq:onsite_EH_measure}
K_{A, {\rm off-diag}}^{n=0} 
= \frac{\beta}{\pi} \int_A  \Psi^\dagger(x) \gamma^0 \Psi(x) dx,
\end{equation}
which is nothing but a mass-like term. 
Recall that all higher-order terms of $n \ge 1$ are in the order of $\mathcal{O}(\beta)$ or higher, therefore at the limit of $\beta \to 0$, we have
\begin{equation}
K_A = \lim_{\beta \to 0} K_{A, {\rm diag}}^{n=0} 
= \int_A \Psi^\dagger(x)  \frac{- i \sigma_3 \partial_x}{z'(x)}  \Psi(x) dx ,
\end{equation}
which reproduces the well-known ground state result. 
It should be noticed that, while we obtain a position-independent onsite potential from the lowest-order perturbation, it could be modified to become position-dependent by considering higher-order terms (see the numerical results in the next Section). 
For the same reason, the position dependence of the first-order derivative in the diagonal component would not be perfectly parabolic as suggested in Eq.~\eqref{eq:NN_EH_measure}.

\section{Numerical results on lattice model}\label{sec:numerics}

We are now ready to discuss the numerical results of the entanglement properties in the lattice model of critical free Dirac fermions under weak measurements. 
To avoid any finite-size effect, in the main text we will focus on the case with an infinite total system size $L \to \infty$, where the two-point correlation functions follow the form of Eq.~\eqref{eq:correlator_lattice}. 
We also numerically calculate the result on large finite total system size $L$, where we find a perfect agreement with the theoretical expectations in the thermodynamic limit (see details in Appendix~\ref{app:numerics}). 
It should be noticed that the high-energy part of the ES of free fermions is very sensitive to numerical errors. 
Its contribution to the EE is small, but is important for investigating the structure of EH. 
To deal with this issue, we use \textit{Mathematica} software to compute the eigenvalue decomposition of the correlation matrix with an extra numerical accuracy (up to hundreds of digits of precision).

In Fig.~\ref{fig:EE_ES}, we present numerical results on the EE and ES for critical free fermions after weak measurements. 
As expected in Eq.~\eqref{eq:defect_effective_central_charge}, both single-interval and two-interval EEs after weak measurements exhibit the same logarithmic scaling behaviors as the ground state case and give a single effective central charge, see Fig.~\ref{fig:EE_ES} (a-c). 
Meanwhile, we find that the ES satisfies the analytically derived relation in Eq.~\eqref{eq:entanglement_spectrum}, which links the spectrum before and after weak measurements. 
The post-measurement ES exhibits a gap $\Delta$ that linearly grows with the measurement strength $\beta$, as shown in Eq.~\eqref{eq:entanglement_spectrum_gap}. 

\begin{figure*}\centering
	\includegraphics[width=\textwidth]{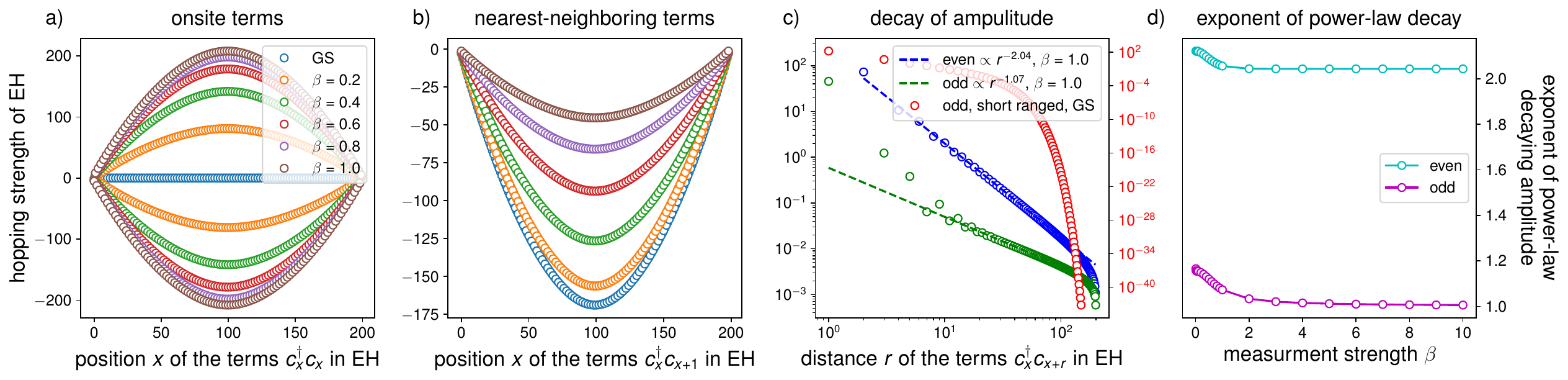}
	\caption{
		\label{fig:EH_decay_main}
		The hopping strength in the EH after weak measurements, calculated with a fixed subsystem size $l=200$. a) The position dependence of onsite terms $c_x^\dagger c_x$ in the EH, exhibiting an even-odd effect (even and odd onsite terms have different signs) as a typical signature of a finite gap. The amplitude of onsite terms monotonically increases with the measurement strength $\beta$ as predicted in Eq.~\eqref{eq:onsite_EH_measure}. b) The position dependence of nearest-neighboring terms $c_x^\dagger c_{x+1}$ in the EH, exhibiting the parabolic behavior of $\frac{1}{z'(x)} \propto x (l - x)$ that is the same as critical ground states. The amplitude of nearest-neighboring terms monotonically decreases with $\beta$ as predicted in Eq.~\eqref{eq:NN_EH_measure}. 
		c) The amplitude of coupling terms $c_x^\dagger c_{x+r}$ in the EH with fixed $\beta = 1$, displaying a power-law decaying with the distance $r$. 
		As expected from our effective field theory, the coupling terms with even distance (blue) and odd distance (green) behave differently with different exponents of the power-law decay. 
		As a comparison, we plot the odd-distance terms of the EH of a critical ground state, as show in red. It decays fast and is small at long-distance to be short-ranged. 
		d) The exponent of power-law decaying couplings, plotted as a function of $\beta$. 
	}
\end{figure*}

To explore the nature of the coexistence of logarithmic EE and gapped ES, we further investigate the structure of EH. 
There are three main features of the EH after weak measurements. 
First, distinct from the critical ground states, there are onsite terms with an even-odd effect (see Fig.~\ref{fig:EH_decay_main} (a)), signaling the presence of a finite gap. 
It is clear to see that the amplitude of onsite terms in EH monotonically increases in terms of the measurement strength $\beta$. 
We numerically find that a fitting of onsite amplitude $\propto \tanh 2 \beta$ works well (see Appendix~\ref{app:numerics}), which gives a linear dependence on $\beta$ for $\beta \ll 1$. 
This is consistent with Eq.~\eqref{eq:onsite_EH_measure} that mass-like onsite terms appear after weak measurements, although from our effective field theory there is no position dependence (other than the even-odd effect) in onsite terms.

Second, as shown in Fig.~\ref{fig:EH_decay_main} (b), the nearest-neighboring hopping exhibits the same parabolic dependence on the position as the critical ground states that are described by CFT, which is also obtained from our effective field theory. 
Its amplitude exhibits a monotonic decrease with increasing $\beta$. 
Moreover, we find that the numerical results suggest a scaling of $\propto \cosh^{-1} 2\beta$. See Appendix~\ref{app:numerics} for more details. 
This is qualitatively consistent with the field-theory calculation, where we obtain an approximated prefactor of $\propto\beta \sinh^{-1} 2\beta$ that is a monotonically decreasing function of $\beta$.

Third and the most important, as shown in Fig.~\ref{fig:EH_decay_main} (c), for any nonzero $\beta$, the EH becomes long-ranged. 
To be specific, the largest amplitude of the hopping terms $c_x^\dagger c_{x + r}$ in the EH exhibits a power-law decay with respect to the distance $r$ between two sites. 
This is in line with our field-theory calculation, which suggests an effective theory with an infinite series of higher-derivatives that could lead to long-range couplings after lattice discretization. 
In contrast, for the critical ground state before measurements (red dots in Fig.~\ref{fig:EH_decay_main} (c)), the EH is short-ranged and rapidly decays with increasing the distance. 
Moreover, we extract the exponent of power-law decaying amplitude of the hopping terms with respect to the distance, and plot it as a function of measurement strength $\beta$ as shown in Fig.~\ref{fig:EH_decay_main} (d). 
Interestingly, we find that there is no sharp change of the exponent. 
For all nonzero $\beta$, the numerical extracted exponents take a stable value of $2$ for even-distance hopping and $1$ for odd-distance hopping. 

\section{Summary and discussion}

Motivated by previous findings on the logarithmic EE of critical ground states under certain measurement protocols~\cite{JianCM2023_measure_ising, JianSK2023_measure_LL, Alicea2024_teleport_critical}, in this work, we have investigated the effect of weak measurement on the entanglement structure of the critical ground state of free Dirac fermions. Importantly, we found that the post-measurement EH is gapped and long-ranged, indicating that conformal symmetry is lost and the state is no longer critical.
Through a field-theory description at continuum, we analytically derived the mapping of the EH before and after measurements, as shown in Eq.~\eqref{eq:int_rep_ker_EH}. 
The post-measurement ES is connected to the ES of a critical ground state via a combination of rescaling and shifting within a hyperbolic scale, as shown in Eq.~\eqref{eq:entanglement_spectrum}, resulting in a gap that linearly increases with the measurement strength $\beta$. 
Meanwhile, the spinor eigenfunction of the EH is modified to reflect the coupling between chiral and anti-chiral components, but its trace is unchanged, as shown in Eq.~\eqref{eq:eigenvector_CM}~\&~\eqref{eq:trace_invariant}. 
This trace is responsible for the finite-size scaling behavior of the reduced density matrix, and its invariance guarantees the presence of a logarithmic EE, similar to that of the critical ground states before measurements.

A closed form of the post-measurement EH is difficult to be obtained, since both the eigenvalues and eigenfunctions of the EH are modified to be in complicated forms. 
Nevertheless, by considering a low-energy expansion, we provide an effective theory that is able to capture several main features observed in the lattice model, including the presence of mass-like onsite terms, the reduction of nearest-neighbor hopping strength, and the emergence of long-range couplings. 
It should be pointed out that the coexistence of a gapped EH and logarithmic EE is unusual. 
Here, as nonunitary operations generally lead to nonlocal effects, the EH becomes long-ranged to make the coexistence possible. 
In our effective theory, the appearance of a series of infinitely higher-order derivatives in the kernel of EH also supports such nonlocal behavior. 
While we do not achieve a closed form, numerically we find that these long-range couplings decay polynomially with distance and behave in a qualitatively different way from the critical ground states.

To better understand our observations of the post-measurement EH, it is beneficial to revisit the known results on critical ground states. 
In the CFT calculation, the single-interval EH of critical ground states is a local integral of the physical Hamiltonian density, i.e., it is local in the real space. 
However, in lattice models there are always long-distance couplings due to the imperfectly linear dispersion relation, for which the CFT description only serves as a low-energy effective theory. 
Taking the free fermionic chain in Eq.~\eqref{eq:massless_fermion_chain} as an example, its dispersion relation is given by $E(k) \propto - \cos (k + k_F)$, where $k_F$ is the Fermi level. 
At half-filling, the low-energy approximation yields $E(k) \propto - \cos (k + \frac{\pi}{2}) = \sin k \sim k$, displaying a linear dispersion. 
This leads to a CFT description of the ground state, for which the EH contains only a first-order derivative term of the field operator as shown in Eq.~\eqref{eq:entanglement_Hamiltonian_CFT_GS}. 
After the lattice discretization, it corresponds to the nearest-neighboring hopping terms as
$K_A^{\rm GS} \propto \sum_{m = 1}^{L} \frac{m (m - L)}{L} c_i^\dagger c_{i+1} + h.c.$. 
On the one hand, it is straightforward to check that the spectrum of the above EH fits well with the ES obtained from the lattice model.
On the other hand, since the CFT description serves only as a low-energy effective theory, the explicit form of the EH on lattice would have a more complicated structure. 
As addressed in Ref.~\cite{Eisler_2017, Eisler2019}, on lattice there are odd-distance hopping terms in the EH that displays in a hyperbolic function of the distance $d$ and subsystem size $l$. 
For short distances $d \ll l$, the hopping amplitude decays polynomially; but for larger distances, it decays quickly to make the EH behaves short-ranged. This is in line with our numerical results shown in Fig.~\ref{fig:EH_decay_main} (c). Physically, this means that truncating the long-distance terms in the EH will not significantly influence the main features of the critical state. We numerically confirmed that this is the case. 
In particular, we find that the logarithmic scaling behavior of EE is robust under the truncation of hopping terms in the EH up to only a few lattice spacing.

In contrast to the critical ground state discussed above, the post-measurement EH displays long-range couplings that are comparable to the subsystem size. 
Moreover, we have numerically verified that truncating the long-range couplings in the post-measurement EH significantly modify the EE and eliminates the logarithmic scaling at large subsystem sizes. 
This is a strong evidence that a long-ranged EH is intrinsic for the critical ground state after weak measurements. 
In addition, the post-measurement EH also hosts long-range couplings at even distances, exhibiting an even-odd effect with a parabolic-like position dependence or a double-peak structure (see Fig.~\ref{fig_app:EH_long_even} in Appendix). 
As we have discussed in the field-theory calculation, these even-distance hopping terms correspond to the coupling between chiral and anti-chiral components. 
This is distinct from the ground state of the Hamiltonian in Eq.~\eqref{eq:massless_fermion_chain} at non-half fillings~\cite{Eisler_2017}, where the hopping terms in EH display a similar position dependence but without an even-odd effect.

Notably, our measurement set-up is strongly related to the interface/defect CFT, as addressed in Ref.~\cite{JianCM2023_measure_ising, JianSK2023_measure_LL, Alicea2024_teleport_critical, JianSK_2023_holo_weak_measure}. 
By converting the effect of weak measurement to an imaginary time evolution driven by a relevant operator (as shown in Eq.~\eqref{eq:imag_time_evo_lattice}), the (path integral representation of the) post-measurement state can be described by a CFT with a defect line at a fixed time. 
When calculating the half-cut EE, a spacetime rotation is allowed to translate the problem to the ground state of the critical free fermionic Hamiltonian with a (partially transmissible) conformal interface located in the middle of the chain. 
For the latter, both the ES and EE for a half-cut along the interface are known~\cite{Peschel2010_EE_defect}, which leads to an analytical understanding to the entanglement structure of the post-measurement states as discussed in Ref.~\cite{JianCM2023_measure_ising}. 
Nevertheless, the spacetime rotation can only be applied to evaluating the trace of reduced density matrix instead of itself, and the EH cannot be achieved in this simple picture. 
Intuitively, the measurement protocol of Eq.~\eqref{eq:imag_time_evo_lattice} is spatially homogeneous, but the interface theory after spacetime rotation is not, which makes the two theories distinguishable and naturally to have different structures of EH. 
It would be interesting to study the EH for the interface theory and make a comparison to our present result on the measurement case.

In the present work we have shown that the effect of weak measurement on the EH can be separated into two parts -- a basis transformation of the eigenvector of the kernel of EH and a modification on the ES. 
In particular, the basis transformation is nontrivial in the following sense. 
On the one hand, the chiral and anti-chiral components of the free Dirac fermion are coupled by this basis transformation. Combining with the fact that a gap is opened in the ES, this leads to a gapped EH with nonlocal couplings that is distinct from a critical ground state. 
On the other hand, for each eigenvalue $s$, the trace of the spinor eigenvector $\Phi_s(x)$ is unchanged under the transformation and independent on $s$. 
Based on these, the EE is separated to two parts: The first part is contributed from the trace of the eigenvector $\Phi_s(x)$ of the kernel of EH, which controls the scaling behavior with subsystem size; The second part is an integral over the $s$ that fully determined by the dispersion relation of EH, which gives the effective central charge. 
This separation implies that the information about the scaling behavior of EE is stored independently of the dispersion of the EH.
We believe this feature not only holds for the free fermions as studied in this work, but also holds for general CFTs. 
To summarize, our work not only provides a detailed study of the entanglement structure of critical ground state under weak measurements, but also reveals the possibility of designing noncritical states that have the same EE scaling behaviors as the critical states but with gapped ES. 
It would be interesting to explore more about the invariance of critical scaling behavior of EE under general physical processes. 
We leave this for a future study~\cite{Tang2024_invariant_EE}.

\medskip
\begin{acknowledgments}
	This work is supported by a startup at Georgia Institute of Technology.
\end{acknowledgments}

\bibliography{WeakMeasure}
\appendix

\renewcommand\thefigure{S\arabic{figure}}
\setcounter{figure}{0} 

\section{Analytical calculation of the two-point correlation function}

\subsection{Discrete lattice model}\label{app:lattice_solution}

Because the imaginary time evolution in Eq.~\eqref{eq:imag_time_evo_lattice} has an A-B structure, it would be convenient to deal with the lattice model in an A-B lattice fashion as
\begin{equation}\label{eq_app:GS_lattice}
\begin{aligned}
H_0 & = - \frac{1}{2} \sum_{m=1}^{L} c_m^\dagger c_{m+1} + h.c. 
\\ 
& = - \frac{1}{2} \sum_{n=0}^{\frac{L}{2}-1} 
a_n^\dagger b_{n + \frac{1}{2}} 
+ a_n^\dagger b_{n - \frac{1}{2}} 
+ h.c.
\end{aligned}
\end{equation}
Then we consider the Fourier transformation of the resized lattice
\begin{equation}
\widetilde{a}_q = \frac{1}{\sqrt{\frac{L}{2}}} 
\sum_{n=0}^{\frac{L}{2}-1} e^{-i \frac{2\pi q n}{\frac{L}{2}}} a_n , 
\quad 
\widetilde{a}_q^\dagger = \frac{1}{\sqrt{\frac{L}{2}}} 
\sum_{n=0}^{\frac{L}{2}-1} e^{+i \frac{2\pi q n}{\frac{L}{2}}} a_n^\dagger , 
\end{equation}
and the inverse transformation
\begin{equation}
a_n = \frac{1}{\sqrt{\frac{L}{2}}} 
\sum_{q=0}^{\frac{L}{2}-1} e^{+i \frac{2\pi q n}{\frac{L}{2}}} \widetilde{a}_q, 
\quad 
a_n^\dagger = \frac{1}{\sqrt{\frac{L}{2}}} 
\sum_{q=0}^{\frac{L}{2}-1} e^{-i \frac{2\pi q n}{\frac{L}{2}}} \widetilde{a}_q^\dagger. 
\end{equation}
Then the Hamiltonian $H_0$ can be rewritten as
\begin{equation}
\begin{aligned}
H_0 & = - \frac{1}{2} \sum_{q=0}^{\frac{L}{2}-1} (
e^{+i \frac{2\pi q \frac{1}{2}}{\frac{L}{2}}} 
+ e^{+i \frac{2\pi q (-\frac{1}{2})}{\frac{L}{2}}} 
) \widetilde{a}_q^\dagger \widetilde{b}_q + h.c. \\ 
& = \sum_{q=0}^{\frac{L}{2}-1} 
\left(
\begin{matrix}
\widetilde{a}_q^\dagger & \widetilde{b}_q^\dagger
\end{matrix}
\right)
\left( - \cos \frac{2\pi q}{L} \right) 
\left(
\begin{matrix}
0 & 1 \\ 1 & 0
\end{matrix}
\right)
\left(
\begin{matrix}
\widetilde{a}_q \\ \widetilde{b}_q
\end{matrix}
\right) . 
\end{aligned}
\end{equation}
It can be further diagonalized by using
\begin{equation}
\left(
\begin{matrix}
0 & 1 \\ 1 & 0
\end{matrix}
\right) 
= \frac{1}{\sqrt{2}} \left(
\begin{matrix}
-1 & 1 \\ 1 & 1
\end{matrix}
\right) 
\left(
\begin{matrix}
-1 & 0 \\ 0 & 1
\end{matrix}
\right) 
\frac{1}{\sqrt{2}} \left(
\begin{matrix}
-1 & 1 \\ 1 & 1
\end{matrix}
\right) ,
\end{equation}
which gives
\begin{equation}
H_0 = \sum_{q=0}^{\frac{L}{2}-1} 
\left(\begin{matrix}
c_{q,-}^\dagger  & c_{q,+}^\dagger
\end{matrix}\right)
\left(
\begin{matrix}
\cos \frac{2\pi q}{L} & 0 \\
0 & - \cos \frac{2\pi q}{L} 
\end{matrix}
\right)
\left(\begin{matrix}
c_{q,-} \\ c_{q,+}
\end{matrix}\right) ,
\end{equation}
with $c_{q,-}^\dagger = \frac{1}{\sqrt{2}}( - \widetilde{a}_q^\dagger + \widetilde{b}_q^\dagger )$, and $c_{q,+}^\dagger = \frac{1}{\sqrt{2}}(  \widetilde{a}_q^\dagger + \widetilde{b}_q^\dagger )$. 
The ground state wavefunction can be written as
\begin{equation}
| \psi_{\rm GS} \rangle 
= \prod_{q \leq |q_F|} c_{q,+}^\dagger | 0 \rangle 
= \prod_{q \leq |q_F|} 
\frac{1}{\sqrt{2}} (\widetilde{a}_q^\dagger + \widetilde{b}_q^\dagger) | 0 \rangle . 
\end{equation}
For the imaginary time evolution in Eq.~\eqref{eq:imag_time_evo_lattice}, we have the driven Hamiltonian as
\begin{equation}
\begin{aligned}
H_{M} & = \sum_{m=0}^{L-1} (-1)^m c^\dagger_m c_m 
= \sum_{n=0}^{\frac{L}{2}-1} a^\dagger_n a_n - b_{n+\frac{1}{2}}^\dagger b_{n+\frac{1}{2}} 
\\ 
& = \sum_{q=0}^{\frac{L}{2}-1} 
\left(
\begin{matrix}
\widetilde{a}_q^\dagger & \widetilde{b}_q^\dagger
\end{matrix}
\right)
\left(
\begin{matrix}
1 & 0 \\ 0 & -1
\end{matrix}
\right)
\left(
\begin{matrix}
\widetilde{a}_q \\ \widetilde{b}_q
\end{matrix}
\right) . 
\end{aligned}
\end{equation}
Then the time evolved wavefunction becomes
\begin{equation}\label{eq_app:lattice_evo}
\begin{aligned}
| \psi(\beta) \rangle & = U(\beta) | \psi_{\rm GS} \rangle 
= e^{-\beta H_M} | \psi_{\rm GS} \rangle 
\\ 
& = \prod_{q \leq |q_F|} \left[ u_q(\beta) \widetilde{a}_q^\dagger + v_q(\beta) \widetilde{b}_q^\dagger \right] | 0 \rangle,
\end{aligned}
\end{equation}
which can be evaluated by the following differential equation
\begin{equation}
\begin{gathered}
\frac{d}{d\beta} 
\left(
\begin{matrix}
u_q(\beta) \\ v_q(\beta) 
\end{matrix}
\right) 
= - \left(
\begin{matrix}
1 & 0 \\ 0 & -1
\end{matrix}
\right) 
\left(
\begin{matrix}
u_q(\beta) \\ v_q(\beta) 
\end{matrix}
\right), 
\end{gathered}
\end{equation}
with $u_q(0)=v_q(0)=1/\sqrt 2$.
The solution is 
\begin{equation}
u_q(\beta) = \frac{1}{\sqrt{2}} e^{-\beta} , \
v_q(\beta) = \frac{1}{\sqrt{2}} e^{\beta} .
\end{equation}
After normalization, we have
\begin{equation}
\frac{|\psi(\beta)\rangle}{\norm{|\psi(\beta)\rangle}} 
= \frac{1}{\sqrt{ e^{-2\beta} + e^{2\beta} }} 
\prod_{q \leq |q_F|} 
(e^{-\beta} \widetilde{a}_q^\dagger + e^{\beta} \widetilde{b}_q^\dagger) | 0 \rangle . 
\end{equation}
From the above normalized time evolved wavefunction, we can calculate its two-point correlation function as
\begin{equation}
\begin{aligned}
\langle a_m^\dagger a_n \rangle_\beta 
& = \frac{e^{-2\beta}}{ e^{-2\beta} + e^{2\beta} } 
\frac{1}{\frac{L}{2}} \sum_{q \leq |q_F|} 
e^{-i \frac{2\pi q (m-n)}{\frac{L}{2}}} \\
\langle b_{m+\frac{1}{2}}^\dagger b_{n+\frac{1}{2}} \rangle_\beta 
& = \frac{1}{ e^{-2\beta} + e^{2\beta} } 
\frac{1}{\frac{L}{2}} \sum_{q \leq |q_F|} 
e^{-i \frac{2\pi q (m-n)}{\frac{L}{2}}} \\ 
\langle a_m^\dagger b_{n+\frac{1}{2}} \rangle_\beta 
& = \frac{e^{2\beta}}{ e^{-2\beta} + e^{2\beta} }  
\frac{1}{\frac{L}{2}} \sum_{q \leq |q_F|} 
e^{-i \frac{2\pi q (m-(n+\frac{1}{2}))}{\frac{L}{2}}} . 
\end{aligned}
\end{equation}
In the case of half-filling, we have the following closed form in the thermodynamic limit $L \to \infty$:
\begin{equation}\label{eq_app:correlator_lattice}
\langle c_m^\dagger c_n \rangle_\beta = 
\begin{cases}
\frac{1}{2} \frac{ e^{-2\beta} }{ \cosh 2\beta }
\quad & {\rm even} \ n = m,  \\ 
\frac{1}{2} \frac{ e^{2\beta} }{ \cosh 2\beta }
\quad & {\rm odd} \ n = m, \\ 
0  \quad & {\rm even} \ n - m \neq 0 , \\ 
\frac{1}{\cosh 2\beta}
\frac{i}{\pi} (n - m)^{-1} \ &  {\rm odd} \ n - m \neq 0 . \\ 
\end{cases}
\end{equation}

\subsection{Continuum field theory}
\label{app:field_solution}

Similar to the lattice model, here for the continuum field theory we also solve the problem by two steps: 1. Solving the ground state; 2. Solving its imaginary time evolution. 
To be more general, here we consider the massive case, and only take the massless limit in the final result of two-point correlation function. 
For the first step, we have the Lagrangian density of free Dirac fermion as
\begin{equation}\label{eq_app:Lagrangian_free_dirac}
\mathcal{L}_0 = \overline{\Psi}(\mathbf{x}) (i \gamma^\mu \partial_\mu - m) \Psi(\mathbf{x}), 
\end{equation}
where $\mathbf{x} = \{ x, t \}$, $\overline{\Psi} = \Psi^\dagger \beta = \Psi^\dagger \gamma^0$ and the Dirac matrices satisfies 
\begin{equation}
\gamma^\mu \gamma^\nu + \gamma^\nu \gamma^\mu = - 2 g^{\mu\nu} \mathbb I_{2\times 2}, 
\quad 
g^{\mu \nu} = {\rm diag} (-1, +1), 
\end{equation}
with $\mu, \nu = 0, 1$. 
In particular, here we choose the convention: 
\begin{equation}
\gamma^0 = \sigma_2, \gamma^1 = i \sigma_1, \gamma^0 \gamma^1 = \sigma_3 .
\end{equation}
In the discussion below, we will ignore the variable of $\mathbf{x}$ if no confusion is caused. 
The Hamiltonian density is
\begin{equation}
\begin{aligned}
\mathcal{H}_0 & = \frac{\partial \mathcal{L}}{\partial (\partial_0 \Psi) } 
\partial_0 \Psi - \mathcal{L} 
= \overline{\Psi} (- i \gamma^1 \partial_1) \Psi + m \overline{\Psi} \Psi \\
& = \Psi^\dagger (-i \gamma^0 \gamma^1 \partial_1) \Psi + m \Psi^\dagger \gamma^0 \Psi ,
\end{aligned}
\end{equation}
where the first part of the Hamiltonian density $\Psi^\dagger (-i \gamma^0 \gamma^1 \partial_1) \Psi$ is the effective theory of the lattice Hamiltonian in Eq.~\eqref{eq_app:GS_lattice}. 

The Lagrangian density in Eq.~\eqref{eq_app:Lagrangian_free_dirac} gives the following equation of motion 
\begin{equation}
(i \gamma^\mu \partial_\mu - m) \Psi(\mathbf{x}) = 0 ,
\end{equation}
which has the following plane-wave solution after quantization
\begin{equation}\label{eq_app:free_dirac_wave}
\begin{aligned}
\Psi(\mathbf{x}) & = \int \widetilde{dk} 
\left[
\hat{b}^\dagger (\mathbf{p}) u (\mathbf{p}) e^{ -i  \mathbf{p}  \mathbf{x} } 
+ \hat{d}  (\mathbf{p}) v (\mathbf{p}) e^{ i  \mathbf{p}  \mathbf{x} }
\right] ,
\\ 
\overline{\Psi}(\mathbf{x}) & = \int \widetilde{dk} 
\left[
\hat{b} (\mathbf{p}) \overline{u} (\mathbf{p}) e^{ i  \mathbf{p}  \mathbf{x} } 
+ \hat{d}^\dagger  (\mathbf{p}) \overline{v} (\mathbf{p}) e^{ -i  \mathbf{p}  \mathbf{x} }
\right] ,
\end{aligned}
\end{equation}
where 
\begin{equation}
\mathbf{p} = \{ k, \omega \}, \quad 
\widetilde{dk} = \frac{dk}{(2\pi) (2\omega)} , \quad \omega = \sqrt{k^2 + m^2} . 
\end{equation}
and the spinors $u(\mathbf{p})$ and $v(\mathbf{p})$ satisfies
\begin{equation}
\begin{aligned}
u \overline{u} & = - (\gamma^\mu p_\mu + m) 
, \quad 
v \overline{v} = - (\gamma^\mu p_\mu -m) . 
\end{aligned}
\end{equation}

The imaginary time evolution is driven by a pure mass term as
\begin{equation}
\mathcal{H}_M = \beta_0 \overline{\Psi}(\mathbf{x}; \tau) \Psi(\mathbf{x}; \tau) .
\end{equation}
It corresponds to the following Lagrangian density
\begin{equation}
\mathcal{L}_M = \overline{\Psi}(\mathbf{x}; \tau) 
(\gamma^0 \partial_\tau + \beta_0)
\Psi(\mathbf{x}; \tau) , 
\end{equation}
which gives 
\begin{equation}
(\gamma^0 \partial_\tau + \beta_0) \Psi(\mathbf{x}; \tau) 
= 0 
\end{equation}
as the equation of motion under the massive imaginary time evolution. Note that here we have the initial condition of the imaginary-time equation of motion as
\begin{equation}
\Psi(\mathbf{x}; \tau = 0) = \Psi(\mathbf{x}) ,
\end{equation}
where $\Psi(\mathbf{x})$ on the right hand side is the solution of Eq.~\eqref{eq_app:free_dirac_wave}. 
After some straightforward algebra, we have
\begin{equation}
\begin{aligned}
\Psi(\mathbf{x}; \tau) & 
= A
\Bigg[
\left( \begin{matrix}
- i \sinh (\beta_0 \tau)  \\ \cosh (\beta_0 \tau)
\end{matrix} \right)
\Psi_1(\mathbf{x}; 0)
\\ & \qquad 
+ \left( \begin{matrix}
\cosh (\beta_0 \tau) \\ i \sinh (\beta_0 \tau) 
\end{matrix} \right)
\Psi_2(\mathbf{x}; 0)
\Bigg] ,
\end{aligned}
\end{equation}
where $\Psi_1(\mathbf{x}; 0)$ and $\Psi_2(\mathbf{x}; 0)$ are the two components of the solution $\Psi(\mathbf{x}) = \left( \Psi_1(\mathbf{x}), \Psi_2(\mathbf{x}) \right)^T$ in Eq.~\eqref{eq_app:free_dirac_wave}. 
Here the normalization factor $A = \cosh^{-\frac{1}{2}} (2\beta_0 \tau)$ is given by letting
\begin{equation}
|A|^2 \overline{\Psi}(\mathbf{x}; \tau) \Psi(\mathbf{x}; \tau) = \overline{\Psi}(\mathbf{x}; 0) \Psi(\mathbf{x}; 0) .
\end{equation}
For simplicity, below we choose $\beta_0 = 1$ as the time scale of the imaginary time evolution. 

Now we are ready to calculate the equal-time propagator as 
\begin{equation}
S_{a,b}(x, x'; \tau) 
= \langle 0|
\Psi_a (x; \tau)  
\overline{\Psi}_b (x'; \tau)  
|0\rangle ,
\end{equation}
where we have taken $t = 0$ and simplified the notation of $\Psi$ to be only dependent on spatial position $x$ instead of $\mathbf{x} = \{x, t\}$. 
After some straightforward algebra, we have
\begin{equation}\label{eq_app:mass_weak_measure_correlator}
\begin{aligned}
& S_{a,b}(x, x'; \tau) 
= \int \widetilde{dk} 
e^{-i \mathbf{p} (\mathbf{x} - \mathbf{x}')} 
\Big[ - m - \omega \tanh 2\tau 
\\ & \quad 
+ (- m \tanh 2 \tau + \omega) \gamma^0 + (- k \cosh^{-1} 2\tau) \gamma^1\Big]_{a, b} .
\end{aligned}
\end{equation}
It reduces to the ordinary well-known result of free Dirac field by taking $\tau = 0$, as
\begin{equation}
S(x, x') = \int \widetilde{dk} 
e^{-i \mathbf{p} (\mathbf{x} - \mathbf{x}')} 
(- \gamma^\mu p_\mu - m) . 
\end{equation}
By taking the massless limit $m \to 0$, Eq.~\eqref{eq_app:mass_weak_measure_correlator} gives
\begin{equation}
\begin{aligned}
\lim_{m \to 0} S(x, x'; \tau) 
& = \frac{1}{2} (-\tanh 2\tau +\gamma^0) \delta(r - r') 
\\ & \quad
- \frac{i}{2\pi} \cosh^{-1} 2\tau \frac{1}{r -r'} \gamma^1. 
\end{aligned}
\end{equation}
For the two-point correlation function we have
\begin{equation}
\begin{aligned}
& C(x, x'; \tau) = \langle 0 | \Psi(x, \tau) \Psi^\dagger(y, \tau) | 0 \rangle 
= S(x, x'; \tau) \gamma^0 \\
= & \frac{1}{2} (1 - \tanh 2\beta \gamma^0) \delta(x - x') 
- \frac{i}{2\pi} \frac{1}{\cosh 2\beta} \frac{\gamma^1 \gamma^0}{x - x'} ,
\end{aligned}
\end{equation}
which is Eq.~\eqref{eq:two_point_field} in the main text. 
Moreover, by taking the explicit matrix form and performing diagonalization, we have the following eigenvalues of the spinor correlation function
\begin{equation}
\lambda = \frac{1}{2} \left( 1 \pm \tanh 2\tau \right) \delta(x-x') 
\pm \frac{i}{2\pi} \frac{1}{\cosh 2\tau} \frac{1}{x-x'} ,
\end{equation}
which is consistent with the lattice result in Eq.~\eqref{eq_app:correlator_lattice}.

\section{Equivalence between the present result of effective central charge and the previous result}

In the main text, we present an integral representation of the effective central charge after weak measurements. 
Here we show that our representation is equivalent to the previous result in Ref.~\cite{Alicea2024_teleport_critical}, which has a different integral representation. 

The entanglement entropy is calculated from the correlation matrix, which has the following eigenvalues
\begin{equation}
\eta_s = \frac{1}{2} \pm \frac{1}{2} \tanh \pi s
= \frac{1}{2} \pm \frac{1}{2} \sqrt{ 
	\left( \frac{\tanh \pi s_0}{\cosh 2\beta} \right)^2 
	+ \left( \tanh 2\beta \right)^2 } .
\end{equation}
Its relation to the prefactor of the logarithmic entanglement entropy is
\begin{equation}\label{eq_app:eff_c_measure}
\begin{aligned}
c^{(n)}_{\rm eff} & = \frac{6n}{1+n} \sigma^{(n)}(\beta) = \frac{12n}{\pi(1+n)} \int ds g^{(n)}(\eta_s) 
\\ & 
= \frac{12n}{\pi(1+n)} \int_{-\infty}^{\infty} d s_0 \frac{ds}{ds_0} g^{(n)}(\eta_{s}) ,
\end{aligned}
\end{equation}
where $g^{(1)}(u) = - u \log u - (1 - u) \log (1 - u)$, 
$g^{(n \neq 1)}(u) = \frac{1}{1-n} \log \left[ u^n + (1-u)^n \right]$.

In (Eq.~(52), (53) in) Ref.~\cite{Alicea2024_teleport_critical}, the authors obtained the following expression of the effective central charge~\footnote{The Ref.~\cite{Alicea2024_teleport_critical} studies the Ising model with central charge $\frac{1}{2}$. To make a comparison with our Dirac fermion result with central charge $1$, in the expression here we have multiplied an additional factor of $2$.}
\begin{equation}
\begin{aligned}
\widetilde{c}^{(n)}_{\rm eff} 
& = \frac{12 n}{\pi^2 (1+n)} \int_{\tanh 2\beta}^1 
f^{(n)}(\lambda) 
\\ & \quad \times 
\log \frac{\sqrt{1-\lambda^2}}{\sqrt{\lambda^2 - \tanh^2 2\beta} + \sech 2\beta} d \lambda, \\
f^{(n)}(\lambda) & = \frac{n}{1-n} \frac{
	(\lambda + 1) (1-\lambda)^n + ( \lambda - 1) (1+\lambda)^n}{
	(\lambda^2 - 1) \left[ (1 - \lambda)^n + (1 + \lambda)^n \right] 
} ,
\end{aligned}
\end{equation}
where $\lambda$ is defined as the eigenvalue of the correlation matrix $\Gamma = \langle \gamma_i \gamma_j \rangle$ of Majorana fermions. It is connected to the entanglement spectrum as $\lambda = \tanh \pi s$.

To see how this expression is related to ours, it is important to notice the following facts. First, 
\begin{equation}
\begin{gathered}
f^{(n)}(\lambda) = \frac{d}{d \lambda} 
\log \left[ (\frac{1+\lambda}{2})^n + (\frac{1-\lambda}{2})^n \right]
= \frac{d}{d \lambda} g^{(n)}( \eta_s(\lambda) ) , 
\\
\eta_s(\lambda) = \frac{1}{2} + \frac{1}{2} \lambda = \frac{1}{2} + \frac{1}{2} \tanh \pi s .
\end{gathered}
\end{equation}
This fact gives us a hint that the previous result in Ref.~\cite{Alicea2024_teleport_critical} and ours are related by an integration by parts, i.e.
\begin{equation}
\begin{aligned}
& \quad 
\int \frac{d}{d \lambda} g^{(n)}( \eta_s(\lambda) ) h(\lambda) d\lambda 
\\ & 
= g^{(n)}( \eta_s(\lambda) ) h(\lambda) 
- \int g^{(n)}( \eta_s(\lambda) ) \frac{d}{d \lambda} h(\lambda) d\lambda . 
\end{aligned}
\end{equation}
Second, we have
\begin{equation}
\begin{aligned}
h(\lambda) & = \log \frac{\sqrt{1-\lambda^2}}{\sqrt{\lambda^2 - \tanh^2 2\beta} + \sech 2\beta} 
\\ & 
= \frac{1}{2} \log \frac{1 - \lambda_0}{1 + \lambda_0} 
= - \arctanh \lambda = - \pi s_0 ,
\end{aligned}
\end{equation}
which means that the function of $\frac{d h (\lambda)}{d\lambda}$ is just a factor of changing integration variable from $s_0$ to $\lambda$. 
It is also obvious that $h(\lambda)$ vanishes for both the upper and lower bounds of the integration interval. 

Based on the above facts, we have 
\begin{equation}
\begin{aligned}
\widetilde{c}^{(n)}_{\rm eff}  
& = - \frac{12n}{\pi^2 (1+n)} \int_{\tanh 2\beta}^{1} 
g^{(n)} ( \eta(\lambda) ) \frac{d s_0}{d\lambda} \frac{dh(\lambda)}{d s_0} d\lambda 
\\ 
& = - \frac{6n}{\pi^2 (1+n)} \int_{-\infty}^{\infty}
g^{(n)} ( \eta(s_0) ) (-\pi) ds_0 
\\ & 
= \frac{6n}{\pi (1+n)} \int_{-\infty}^{\infty}
g^{(n)} ( \eta(s_0) ) ds_0 ,
\end{aligned}
\end{equation}
which is identical to our expression of Eq.~\eqref{eq_app:eff_c_measure} upto to a factor of $2$ that comes from the difference between Ising model (with central charge $\frac{1}{2}$) and Dirac fermion (with central charge $1$). Here in the second line we have used the fact that $g^{(n)}(\eta(s_0)) = g^{(n)}(\eta(-s_0))$.

\section{More about the numerical calculation on the lattice model}\label{app:numerics}

The numerical calculations in the present work are performed in a two-fold way. 
On the one hand, for the calculation of two-point correlation functions and entanglement entropy, it is safe to use the ordinary floats to do the linear algebra efficiently. 
In these calculations, we can perform a ground-up calculation starting from the Hamiltonian and a fixed filling factor (or equivalently a given chemical potential), without any other prior knowledge.
On the other hand, the values of entanglement spectrum are very close to $0$ and $1$, with the differences between various values being much smaller than ordinary floating-point error. Consequently, the calculation of the entanglement Hamiltonian requires a level of accuracy that is significantly higher than that of standard floating-point calculations.
To address this issue, we use the \textit{Mathematica} package for linear algebra with precision up to hundreds of significant digits. 
Moreover, to reduce computational cost and avoid the accumulation of errors, we start from an analytical solution of the two-point correlation functions on a discrete chain of infinite length, rather than using the Hamiltonian directly. 
It has been checked that the resulting entanglement Hamiltonian produces consistent results for the entanglement entropy when compared to ordinary numerical calculations with standard floating-point precision.

\begin{figure*}\centering
	\includegraphics[width=\textwidth]{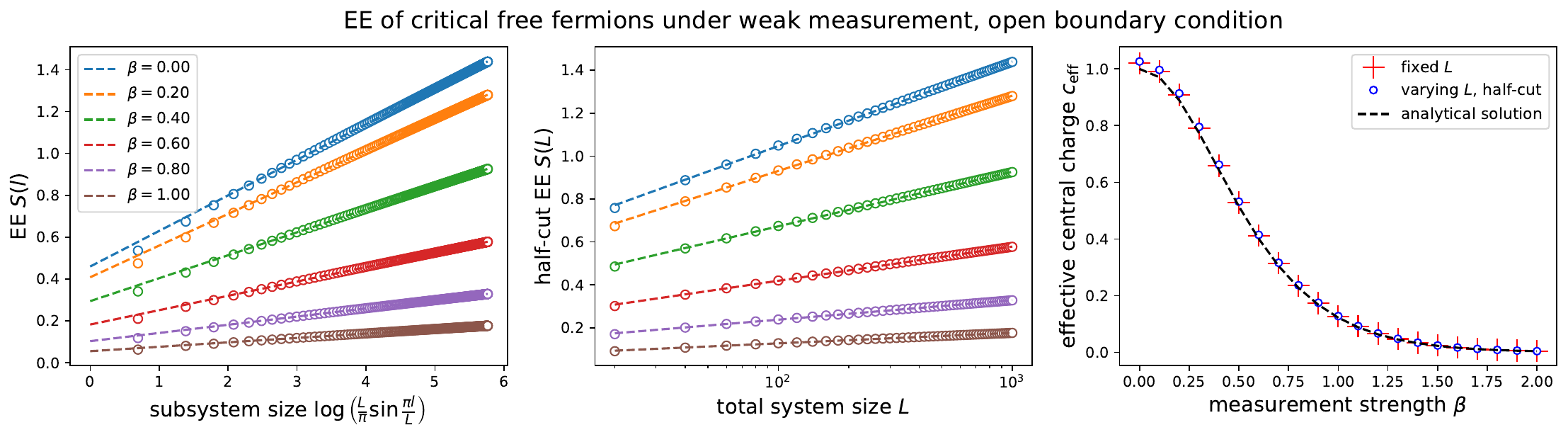}
	\caption{
		\label{fig_app:EE_open_finite_lattice}
		The single-interval EE of critical free fermions before ($\beta =0$) and after ($\beta \neq 0$) weak measurements, calculated from finite chains with open boundary conditions and a largest considered total system size $L = 1000$.
	}
\end{figure*}

\begin{figure*}\centering
	\includegraphics[width=\textwidth]{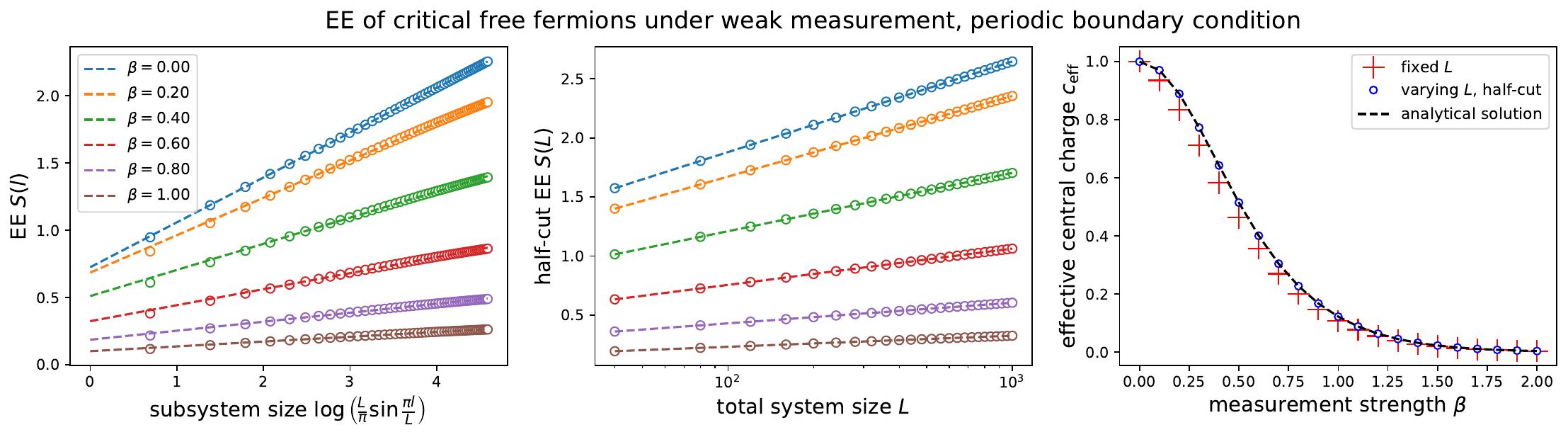}
	\caption{
		\label{fig_app:EE_periodic_finite_lattice}
		The single-interval EE of critical free fermions before ($\beta =0$) and after ($\beta \neq 0$) weak measurements, calculated from finite chains with  periodic boundary conditions and a largest considered total system size $L = 1000$.
	}
\end{figure*}

\begin{figure*}\centering
	\includegraphics[width=0.75\textwidth]{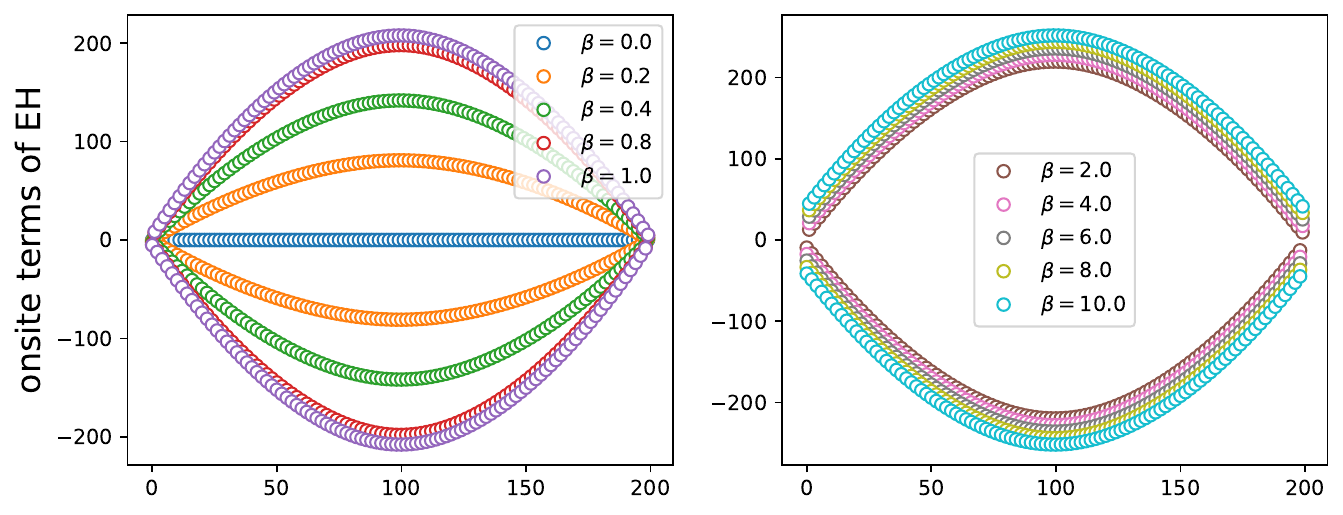}
	\caption{
		\label{fig_app:EH_onsite}
		The strength of onsite terms in the entanglement Hamiltonian before ($\beta = 0$) and after ($\beta \neq 0$) weak measurements. 
		For all nonzero $\beta$, the onsite terms display an even-odd effect as a signature of a finite gap.  
		When the measurement strength $\beta$ is small (left panel), changing $\beta$ only modifies the amplitude. When $\beta$ is large (right panel), there is a gap between even and odd parts. 
		Numerically we find that the spatial dependence of the onsite terms fits good with the parabolic function as the nearest-neighboring term.
		Here the subsystem size is $l = 200$ and the total system size is considered to be infinite.
	}
\end{figure*}

\begin{figure*}\centering
	\includegraphics[width=\textwidth]{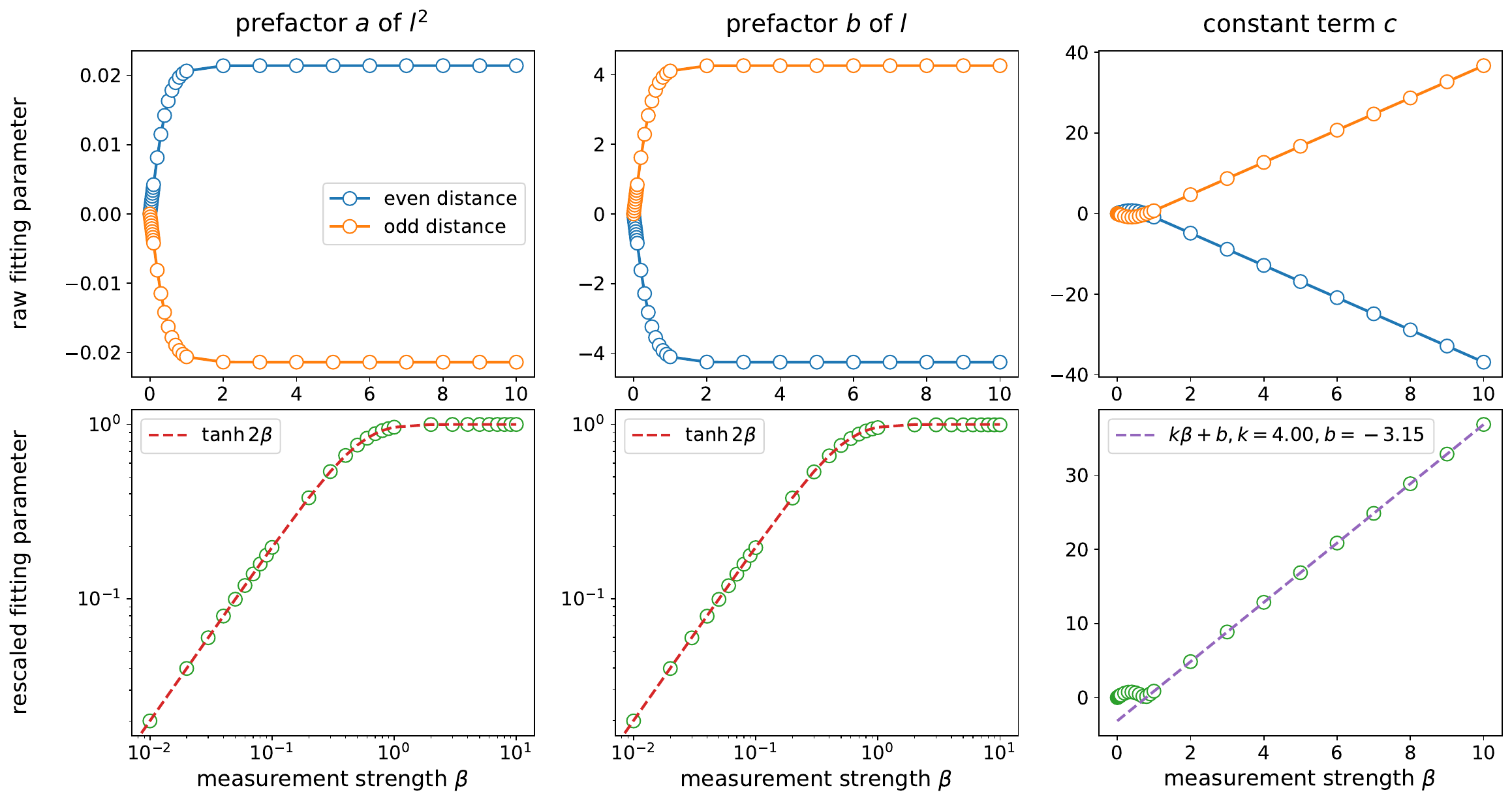}
	\caption{
		\label{fig_app:EH_onsite_fit}
		The fitting parameter of $J_{x, x} = a x^2 + b x + c$, where $J_{x,x}$ is the coupling constant of onsite terms $J_{x,x} c^\dagger_x c_x$ in the EH. 
		Upper panel: the raw fitting data. Lower panel: the rescaled fitting parameter of letting $a = 1$ for the largest considered $\beta$. 
	}
\end{figure*}

\begin{figure*}\centering
	\includegraphics[width=0.75\textwidth]{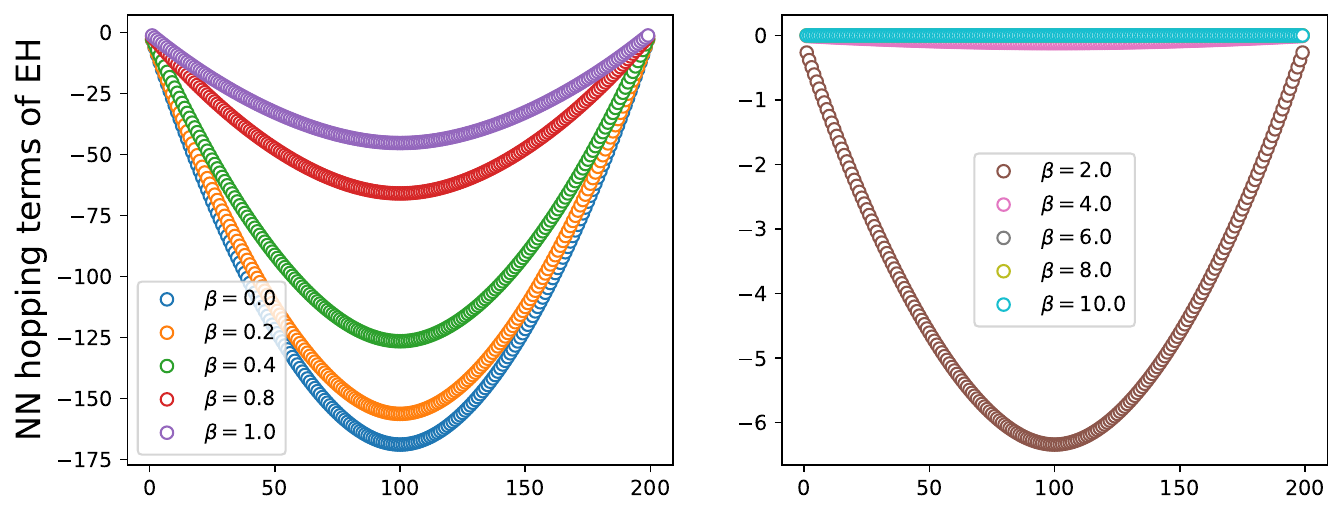}
	\caption{
		\label{fig_app:EH_NN}
		The strength of nearest-neighboring terms in the entanglement Hamiltonian before ($\beta = 0$) and after ($\beta \neq 0$) weak measurements. 
		For all nonzero $\beta$, the onsite terms display an even-odd effect as a signature of a finite gap.  
		When the measurement strength $\beta$ is small (left panel), changing $\beta$ only modifies the amplitude. When $\beta$ is large (right panel), there is a gap between even and odd parts. 
		Here the subsystem size is $l = 200$ and the total system size is considered to be infinite.
	}
\end{figure*}

\begin{figure*}\centering
	\includegraphics[width=\textwidth]{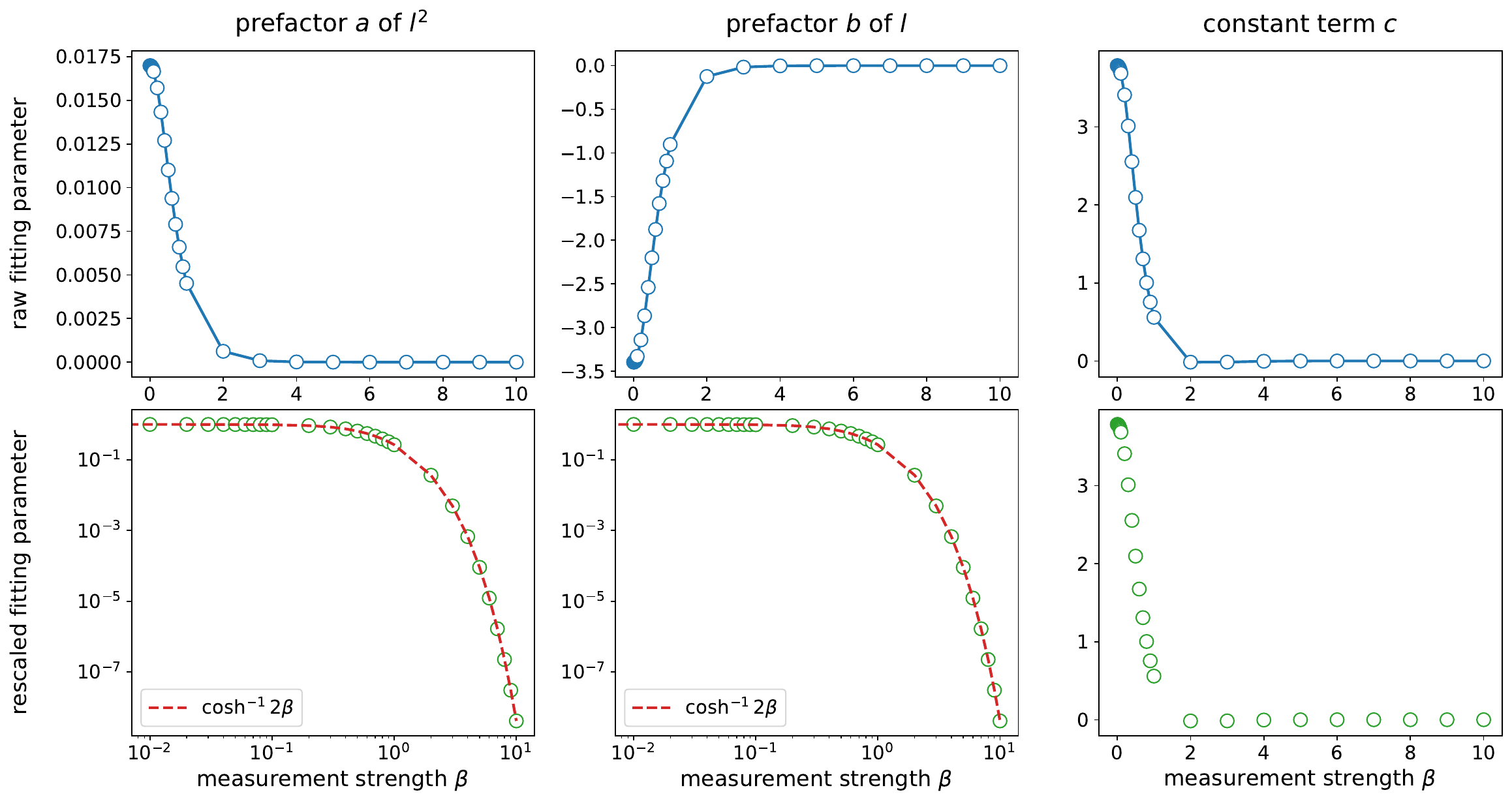}
	\caption{
		\label{fig_app:EH_NN_fit}
		The fitting parameter of $J_{x, x+1} = a x^2 + b x + c$, where $J_{x,x+1}$ is the coupling constant of nearest-neighboring terms $J_{x,x+1} c^\dagger_x c_{x+1} + h.c.$ in the EH. 
		Upper panel: the raw fitting data. Lower panel: the rescaled fitting parameter of letting $a = 1$ for the ground state case $\beta = 0$. 
	}
\end{figure*}

\begin{figure*}\centering
	\includegraphics[width=\textwidth]{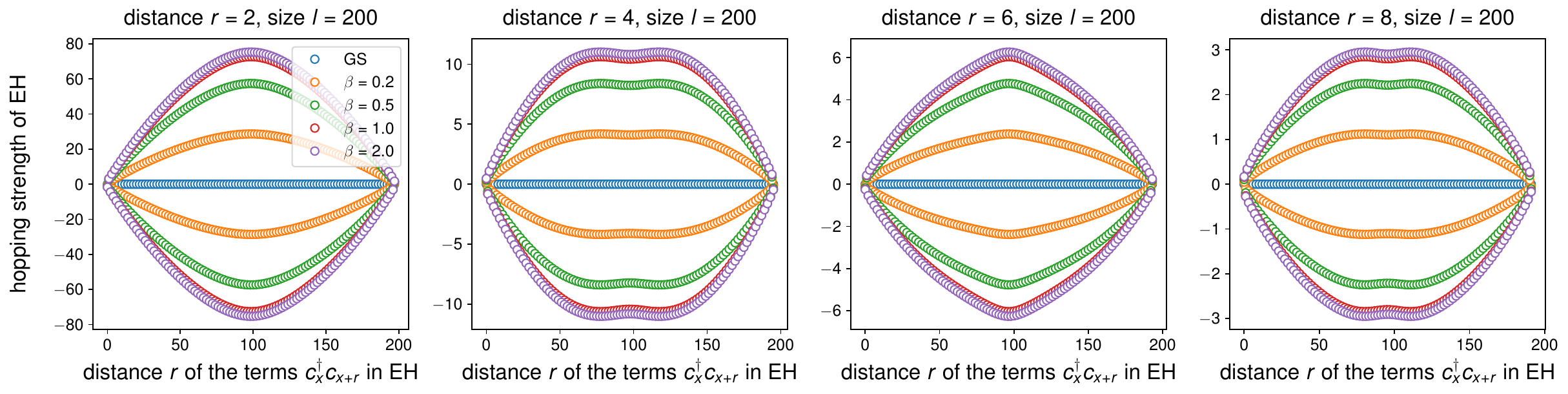}
	\caption{
		\label{fig_app:EH_long_even}
		The strength of the first few even-distance hopping terms in the entanglement Hamiltonian before ($\beta = 0$) and after ($\beta \neq 0$) weak measurements. 
		For all nonzero $\beta$, the even-distance hopping terms display an even-odd effect. 
		When the distance $r\,\, {\rm mod}\,\, 4 = 0$, it displays a double-peak structure. 
		Here the subsystem size is $l = 200$ and the total system size is considered to be infinite.
	}
\end{figure*}

\begin{figure*}\centering
	\includegraphics[width=\textwidth]{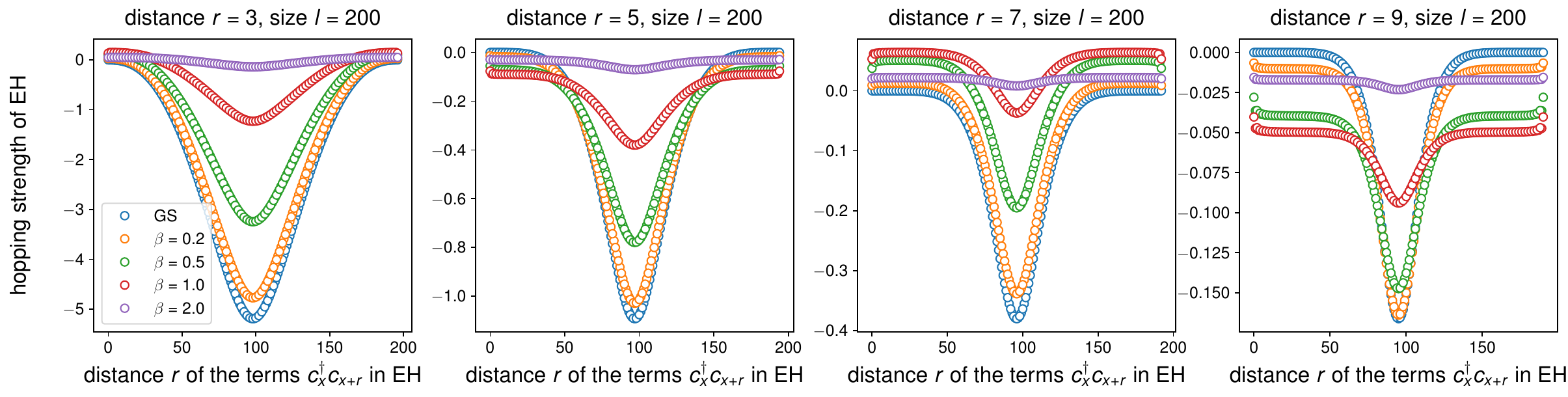}
	\caption{
		\label{fig_app:EH_long_odd}
		The strength of the first few odd-distance hopping terms in the entanglement Hamiltonian before ($\beta = 0$) and after ($\beta \neq 0$) weak measurements. 
		As increasing the distance $r$, the position dependence becomes sharper. 
		Here the subsystem size is $l = 200$ and the total system size is considered to be infinite.
	}
\end{figure*}

In the main text, we mainly focus on the case of an infinite total system size to avoid possible finite-size effect.
Here, we present numerical results of EE on a finite lattice for a comparison. 
As shown in Fig.~\ref{fig_app:EE_open_finite_lattice} and Fig.~\ref{fig_app:EE_periodic_finite_lattice}, both open and periodic boundary conditions gives the expected logarithmic scaling that is identical to the pre-measurement critical ground state, resulting a single effective central charge as discussed in the main text. 
In Fig.~\ref{fig_app:EH_onsite} \&~\ref{fig_app:EH_NN}, we present numerical results for the onsite and nearest-neighbor terms in the entanglement Hamiltonian over a wider range of measurement strengths $\beta$. 
Similar to the ground state of a gapless fermionic chain at non-zero fillings~\cite{Eisler_2017}, both the onsite and nearest-neighboring terms in post-measurement entanglement Hamiltonian exhibit a parabolic dependence on the spatial position. 
For small $\beta$, the change resembles a modification of the amplitude of the parabolic function. For large $\beta$, the amplitude remains stable while the onsite terms shift linearly with respect to the measurement strength $\beta$. 
Through a numerical fitting, we find that the amplitude (the prefactor of the parabolic fit) is proportional to  $\tanh 2\beta$ for the onsite terms and $\cosh^{-1} 2\beta$ for the nearest-neighbor terms, as shown in Fig.~\ref{fig_app:EH_onsite_fit} \&~\ref{fig_app:EH_NN_fit}, respectively. 
Moreover, we also plot the strength of the first few distant hopping terms in the entanglement Hamiltonian in Fig.~\ref{fig_app:EH_long_even} \&~\ref{fig_app:EH_long_odd}.
For even-distance hopping terms, as shown in Fig.~\ref{fig_app:EH_long_even}, we observe a double-peak structure, which was reported for the gapless ground state at non-half filling~\cite{Eisler_2017}. 
We find that the double-peak structure appears when the subsystem size $l \,\,{\rm mod}\, 4 = 0$. 
For odd-distance hopping terms, as shown in Fig.~\ref{fig_app:EH_long_odd}, the spatial position dependence is flatten as we increase the measurement strength $\beta$, and becomes sharper as we increase the hopping distance $r$.

\begin{figure*}\centering
	\includegraphics[width=\textwidth]{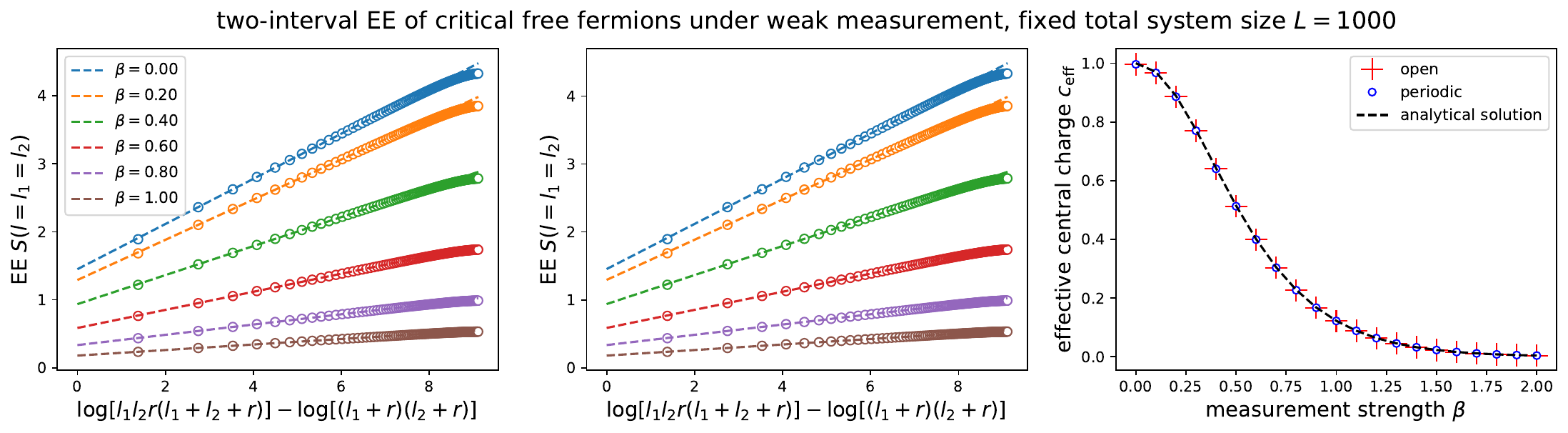}
	\caption{
		\label{fig_app:TwoInt_EE_finite_latice}
		The two-interval EE of critical free fermions before ($\beta =0$) and after ($\beta \neq 0$) weak measurements, calculated from finite chains with (left) open and (middle) periodic boundary conditions. The total system size is fixed to be $L = 1000$ and the distance between two intervals is chosen to be $r = 21$. 
	}
\end{figure*}

\clearpage

\end{document}